\documentclass[english,12pt,aps,prd,a4paper,preprintnumbers,floatfix,nofootinbib,showpacs,superscriptaddress, notitlepage]{revtex4-1} 
\usepackage[mode=buildnew]{standalone}
\usepackage[usenames,dvipsnames]{color}  
\usepackage{graphicx}

\usepackage{setspace}
\usepackage{caption}
\usepackage{subcaption}
\usepackage{amsmath}
\usepackage{amssymb}
\usepackage[colorlinks=true,citecolor=darkred,urlcolor=darkred, pdfborder={0 0 0}]{hyperref}
\usepackage[normalem]{ulem}
\usepackage{xcolor}
\usepackage{placeins}
\usepackage{mathrsfs}
\usepackage{verbatim} 
\usepackage{cancel} 
\usepackage{float} 

\usepackage{scalerel}
\usepackage{tikz-feynman}
\tikzfeynmanset{compat=1.1.0}
\usepackage{feynmp}
\usepackage{tikzsymbols}
\usepackage{multirow}

\makeatletter
\def\p@subsection{}
\makeatother

\definecolor{darkred}{rgb}{0.6,0,0}

\definecolor{linkcolor}{rgb}{0,0,0.5}

\def\gsim{\raise0.3ex\hbox{$\;>$\kern-0.75em\raise-1.1ex\hbox{$\sim\;$}}}
\def\lsim{\raise0.3ex\hbox{$\;<$\kern-0.75em\raise-1.1ex\hbox{$\sim\;$}}}

\def\beqn#1{\begin{equation}\label{#1}}
\def\eeqn{\end{equation}}

\def\beqa#1{\begin{eqnarray}\label{#1}}
\def\eeqa{\end{eqnarray}}

\newcommand {\ignore}[1]{}

\def\Z4{$Z_4$}
\def\O5{$\mathcal{O}_5$ }

\def\321{$\mathrm{SU(3) \otimes SU(2) \otimes U(1)}$ }

\usepackage{orcidlink}
 
 \newcommand{\AddrIISERB}{Department of Physics,
 Indian Institute of Science Education and Research - Bhopal \\
 Bhopal Bypass Road, Bhauri, Bhopal, India}

\begin{document}

\title{\color{BrickRed} Minimal $A_4$ Type-II Seesaw Realization of Testable Neutrino Mass Sum Rules}

 \author{Salvador Centelles Chuliá\orcidlink{0000-0001-6579-1067}}\email{salcen@ific.uv.es}
 \affiliation{Instituto de Fısica Corpuscular, CSIC-Universitat de Valencia, 46980 Paterna, Spain}
\author{Ranjeet Kumar\orcidlink{0000-0002-7144-7606}}\email{kumarranjeet.drk@gmail.com}
\affiliation{Institute for Convergence of Basic Studies, Seoul National University of Science and Technology, Seoul 01811, Republic of Korea}
\affiliation{\AddrIISERB}

\begin{abstract}
  \vspace{1cm} 
We propose a flavour model based on an $A_4$ symmetry combined with a type-II seesaw mechanism for neutrino mass generation. The resulting neutrino mass matrix obeys a sum rule that, together with the measured mass-squared differences, fully determines the absolute neutrino mass spectrum. The constrained flavour structure yields correlated predictions for lepton mixing parameters, leads to inverted ordering after imposing mixing constraints, restricts the Majorana phases and implies a neutrinoless double beta decay rate close to its maximal value for inverted ordering. In the charged lepton sector an approximate triality symmetry arises in the seesaw limit, suppressing muon flavour-violating processes and allowing only specific $\tau$ decay channels. The model provides a tightly constrained and experimentally testable framework linking neutrino masses, lepton mixing and lepton-number-violating observables.
\end{abstract}

\maketitle


\section{Introduction} \label{sec:intro}

One of the open questions in the Standard Model (SM) is the pattern of fermion masses and mixings, i.e.\ the \textit{flavour puzzle}. Indeed, the fermion mass pattern and quark mixing are not predicted by the theory: the Standard Model fits the observed fermion masses through Yukawa couplings, while lacking neutrino masses altogether at the renormalizable level. Discrete symmetries provide a powerful theoretical framework for alleviating the flavour puzzle by reducing the number of free parameters in the theory \cite{Babu:1990fr, Frampton:1994rk, Kubo:2003iw, Altarelli:2010gt, Morisi:2012fg, Feruglio:2019ybq, Denton:2023hkx, Ding:2024ozt}. Symmetries such as the tetrahedral group $A_4$ \cite{Ishimori:2010au} have been widely studied over the last two decades; see \cite{Babu:2002dz,Chen:2005jm,Altarelli:2005yx,CarcamoHernandez:2013yiy,CarcamoHernandez:2015rmj,Borah:2017dmk,CarcamoHernandez:2017kra,CentellesChulia:2017koy,Borah:2018nvu,Ding:2020vud,Devi:2022scm,Bora:2022jdp,RickyDevi:2023fqd,Kumar:2023moh,Mahapatra:2023oyh,Singh:2024imk,Kumar:2024zfb,Nomura:2024atp,Palavric:2024gvu,Pathak:2024sei,Goswami:2025jde,Moreno-Sanchez:2025bzz,Kumar:2025cte,Kumar:2025zvv} for a selection of examples. Additionally, the breaking pattern $A_4 \to Z_3$ also leads to lepton triality~\cite{Ma:2001dn,Ma:2010gs,Cao:2011df}, which typically imposes strong constraints on the possible signals of charged lepton flavour violation (cLFV). See \cite{Calibbi:2025fzi} for a generalization of the triality residual symmetry and \cite{deMedeirosVarzielas:2025byb} for an analysis on residual symmetries after spontaneous breaking of a non-abelian symmetry.

Despite the phenomenological success of flavour symmetries in reproducing fermion mixing patterns, the mechanism responsible for neutrino mass generation remains unresolved. The discovery of neutrino oscillations~\cite{Kajita:2016cak,McDonald:2016ixn} provides clear evidence for physics beyond the Standard Model, most straightforwardly interpreted as the existence of non-zero neutrino masses. Among the proposed explanations, the seesaw paradigm offers a particularly compelling framework for understanding their smallness. In this work, we focus on the type-II seesaw mechanism~\cite{Schechter:1980gr,Magg:1980ut,Cheng:1980qt,Mohapatra:1980yp}, which does not require the introduction of new fermionic degrees of freedom beyond the Standard Model. Neutrino masses are instead generated through the vacuum expectation value of an $SU(2)_L$ scalar triplet $\Delta$. Furthermore, the type-II seesaw can also account for the generation of the baryon asymmetry of the Universe via leptogenesis~\cite{Ma:1998dx,Antusch:2004xy,Antusch:2007km,Barrie:2021mwi,Barrie:2022cub,Berbig:2025hlc}. When combined with flavour symmetries, this framework becomes especially appealing, as it can simultaneously address the origin of neutrino masses and the observed lepton mixing patterns. The phenomenology of the type-II seesaw has been studied extensively in the literature; see, for instance, \cite{Lavoura:2003xp,Dinh:2012bp,Barrie:2022ake,Mandal:2022zmy}.

Another interesting aspect of flavour symmetries is the realization of neutrino mass ``sum rules''. In their traditional form, these provide relations among the complex eigenvalues of the neutrino mass matrix~\cite{Barry:2010yk,King:2013psa,Gehrlein:2016wlc,Gehrlein:2020jnr}. More recently, mass sum rules have been derived for the singular values of the mass matrix~\cite{ChuliaCentelles:2022ogm,CentellesChulia:2023osj}, which correspond directly to the physical neutrino masses rather than to its complex eigenvalues. In this framework, the experimental determination of the neutrino mass-squared differences uniquely fixes the absolute neutrino mass scale. As a consequence, the allowed parameter space relevant for beta decay and neutrinoless double beta decay (\(0\nu ee\)) processes is significantly constrained. This approach therefore provides a complementary probe, alongside neutrino oscillation experiments, for distinguishing among different flavour models.

In this work, we develop a simple and predictive model based on the flavour symmetry $A_4$ that leads to a neutrino mass sum rule and an approximate lepton triality in the charged lepton sector in the seesaw limit. Neutrino masses are generated via the type-II seesaw mechanism by introducing an $SU(2)_L$ scalar triplet $\Delta$, which also transforms as a triplet under $A_4$. An $SU(2)_L$ doublet $H$ is likewise assigned to transform as a triplet under $A_4$, with its vacuum alignment $(1,1,1)^T$ spontaneously breaking the flavour symmetry down to its $Z_3$ subgroup, giving rise to lepton triality as a residual symmetry of the charged lepton sector in the seesaw limit $m_\Delta \gg \Lambda_{EW}$.

As a consequence of this structure, the charged lepton mass matrix is diagonalized by the so-called magic matrix, up to a permutation matrix. Combining the charged lepton and neutrino mixing matrices, we obtain a robust and testable correlation between the atmospheric mixing angle $\theta_{23}$ and the Dirac CP-violating phase $\delta_{\rm CP}$. This correlation represents a key prediction of the model and can be probed in current and future neutrino oscillation experiments. In addition, the neutrino mass sum rule fixes the absolute neutrino mass scale once the measured mass-squared differences are imposed, leading to sharp predictions for neutrinoless double beta decay and beta decay observables. Charged lepton flavour violating processes are constrained by the approximate lepton triality.

The remainder of the paper is organized as follows. In Sec.~\ref{sec:model}, we present the field content, symmetry assignments, and vacuum alignments of the model, as well as the charged lepton and neutrino mass matrices. In Sec.~\ref{sec:sumrule}, we derive the neutrino mass sum rule and its phenomenological consequences. In Sec.~\ref{sec:mixing}, we analyze the mixing predictions of the model, including neutrino oscillations, CP violation and lepton number breaking observables such as neutrinoless double beta decay. In Sec.~\ref{sec:LFV} we discuss the cLFV processes allowed by the approximate triality symmetry within this model. Finally, we summarize our results and present our conclusions in Sec.~\ref{sec:conclusions}.

\section{Model and flavour structure} 
\label{sec:model}

We consider a type-II seesaw framework for neutrino masses with a flavour \(A_4\) symmetry which fixes the flavour structure of the lepton sector. The model leads to a neutrino mass sum rule with important implications for a variety of experiments and observations. In addition, the setup yields a rich and predictive phenomenology for cLFV. The scalar sector relevant for leptons consists of an \(SU(2)_L\) Higgs doublet \(H=(H_1,H_2,H_3)\) and an \(SU(2)_L\) triplet \(\Delta=(\Delta_1,\Delta_2,\Delta_3)\), both transforming as \(A_4\) triplets. The Lepton doublet $L$ is also a flavour triplet, while the right handed charged leptons transform as $(1, 1', 1'')$. On the other hand, quark masses require an additional \(SU(2)_L\) doublet \(H_q\) that is a flavour singlet; see Appendix~\ref{app:quarkmass}.
The particle content and their transformation properties under different symmetries are summarized in Table~\ref{tab:fields}. 

\begin{table}[!h]
    \centering
    \begin{tabular}{|c||c|c|c|}
        \hline 
        \ Fields \ &  \ $SU(2)_L$ \ & \ $U(1)_Y$ \ & \ $ \mathbf{A_4}$  \ \\ \hline
         $L$ &   $2$ & $-1/2$ & $\mathbf{3}$  \\
         ${l_{R}}$ &  $1$ & $-1$ & $\mathbf{1, 1'', 1'}$  \\ 
         \hline
         $H$ &  $2$ & $1/2$ & $\mathbf{3}$   \\
         $\Delta$  & $3$ & $1$ & $\mathbf{3}$  
         \\ 
         \hline 
    \end{tabular}
    \caption{Particle content of the lepton sector of the model and their transformation properties under various symmetries.}
    \label{tab:fields}
\end{table}

The field transformations are chosen such that the Higgs vev alignment $v(1,1,1)$ leads to the so-called magic matrix $U_m$, which diagonalizes the charged lepton mass matrix. Additionally, this vev alignment preserves a residual $Z_3$ symmetry after the spontaneous breaking of $A_4$. This residual $Z_3$ symmetry has interesting phenomenological consequences for the LFV as we discuss in Sec.~\ref{sec:LFV}. In the neutrino sector, the vev alignment $\langle \Delta \rangle = (u_1,u_2,u_3)^T$ of the scalar triplet is responsible for the emergence of a neutrino mass sum rules. This sum rule is valid irrespective of the neutrino mass orderings. Once we impose the observed mass-squared differences constraint, this sum rule predicts the precise values for the absolute neutrino masses. Having discussed the model framework, next we explore the neutrino mass generation and mixing.
\subsection{Lagrangian and Mass Matrices} \label{sec:lag}

Following the charge assignment provided in table \ref{tab:fields}, the invariant Yukawa Lagrangian under SM gauge symmetries and $A_4$ can be formulated as
\begin{align} \label{eq:lag1}
    -\mathcal{L}= &\alpha_1 \left( \bar{L} \otimes H \right)_\mathbf{1} \otimes \left(l_{R_1}\right)_{\mathbf{1}} +\alpha_2 \left( \bar{L} \otimes H \right)_{\mathbf{1'}} \otimes \left(l_{R_2}\right)_{\mathbf{1''}} + \alpha_3 \left( \bar{L} \otimes H \right)_{\mathbf{1''}} \otimes \left(l_{R_3}\right)_{\mathbf{1'}} \nonumber \\
    &+\alpha \left( \bar{L^c} \otimes L\right)_{\mathbf{3S}} \otimes \left(i \tau_2 \Delta \right)_{\mathbf{3}} + \text{h.c.} \ ,
\end{align}
where $[...]_\mathbf{p}$; $\mathbf{p = 1, 1',1''
, 3_S, 3}$ denote the $A_4$ transformation of enclosed fields. The above Lagrangian can be further expanded following the $A_4$ tensor product rules and given as
\begin{align} \label{eq:lag2}
    -\mathcal{L}= &\alpha_1 \left( \bar{L}_1 H_1 + \bar{L}_2 H_2 + \bar{L}_3 H_3  \right) l_{R_1} + \alpha_2 \left( \bar{L}_1 H_1 + \omega \bar{L}_2 H_2 + \omega^2 \bar{L}_3 H_3  \right) l_{R_2}  \nonumber \\ &+\alpha_3 \left( \bar{L}_1 H_1 + \omega^2 \bar{L}_2 H_2 + \omega \bar{L}_3 H_3  \right) l_{R_3} \nonumber \\
    &+\alpha \left[ \left( \bar{L}^c_2 L_3 +  \bar{L}^c_3 L_2 \right) i \tau_2 \Delta_1 + \left( \bar{L}^c_3 L_1 +  \bar{L}^c_1 L_3 \right) i \tau_2 \Delta_2 + \left( \bar{L}^c_1 L_2 +  \bar{L}^c_2 L_1 \right) i \tau_2 \Delta_3  \right]+ \text{h.c.} \ .
\end{align}
Once the scalar fields $H$ and $\Delta$ acquire vevs, the charged lepton and neutrino masses are generated. We focus on a particular choice for the vev alignment of these scalars. The vev alignment $\langle H \rangle = v(1,1,1)^T$ preserves the residual $Z_3$ symmetry, which have important implications for LFV processes. The choice of  $\langle \Delta \rangle =  (u_1, u_2,u_3)^T$ leads to the neutrino mass sum rule. From Eq.~\eqref{eq:lag2}, the charged lepton and neutrino mass matrices can be extracted as follows 
\begin{align}
\mathcal{M}_l= \sqrt{3} \, v \, U_m \begin{pmatrix}
    \alpha_1 & 0 & 0 \\
     0 & \alpha_2 & 0 \\
     0 & 0 & \alpha_3 
\end{pmatrix}, \quad
\mathcal{M}_{\nu}= \alpha \begin{pmatrix}
    0 & u_3 & u_2 \\
    u_3 & 0& u_1 \\
     u_2 & u_1 & 0 
\end{pmatrix} \quad .
\end{align}
Where, we have defined the so-called magic matrix $U_m$ as
\begin{equation}
    U_m = \frac{1}{\sqrt{3}} \begin{pmatrix}
    1 & 1 & 1 \\
    1 & \omega & \omega^2 \\
    1 & \omega^2 & \omega 
\end{pmatrix}.
\end{equation}
We can, without loss of generality, rephase the right-handed charged leptons and the flavour triplet $L$ to take the couplings $\alpha_i$ and $\alpha$ real and positive. Additionally, we redefine
\begin{equation}
\langle \Delta \rangle
= u \, e^{i \eta_1}
\bigl(\cos \theta ,\, \sin \theta \cos \phi \, e^{i \eta_2}, \,
\sin \theta \sin \phi \, e^{i \eta_3}\bigr)^T ,  \label{eq:alignment}
\end{equation}
where $u=\sqrt{|u_1|^2+|u_2|^2+|u_3|^2}>0$, the alignment angles $\theta$ and $\phi$ lie in the range $[0,\pi/2]$, such that their trigonometric functions are non-negative and determine the relative alignment of the triplet vev, while the phases $\eta_i$ take values in $[0,2\pi)$.

The charged lepton mass matrix can be diagonalized as
\begin{equation}
    U_L^\dagger \mathcal{M}_\ell U_R = \text{diag}(m_e, m_\mu, m_\tau)\, .
\end{equation}
Since $U_m$ is unitary, up to permutations the only possible solution is
$U_L =  U_m\, P$, with $P$ a permutation matrix, and
$(m_e, m_\mu, m_\tau) = \sqrt{3}\, v\, (|\alpha_i|,\, |\alpha_j|, \, |\alpha_k|)$.
In the neutrino sector, the mass matrix is diagonalized according to
\begin{equation}
    U_\nu^T \mathcal{M}_\nu U_\nu = \text{diag}(m_1, m_2, m_3)\, ,
    \label{eq:diagnu}
\end{equation}
so that physical lepton mixing arises from the mismatch between the diagonalization matrices of the charged lepton and neutrino sectors and is encoded in the lepton mixing matrix, $U_{\ell}$, as
\begin{equation}
    U_{\ell}= U_L^{\dagger} U_{\nu}\, .
\end{equation}

We choose the symmetric parametrization of the lepton mixing matrix \cite{Schechter:1980gr, Rodejohann:2011vc}

\begin{equation}
U_{\ell}= P(\delta_1, \delta_2, \delta_3) \, U_{23}(\theta_{23}, \phi_{23}) \, U_{13} (\theta_{13}, \phi_{13}) \, U_{12}(\theta_{12}, \phi_{12}) \, ,
\end{equation}
where $P(\delta_1, \delta_2, \delta_3)$ is a diagonal matrix of unphysical phases and the $U_{ij}$ are complex rotations in the $ij$ plane, as for example,
\begin{equation}
    U_{23} (\theta_{23}, \phi_{23}) = \left(\begin{matrix}
        1 & 0 & 0 \\
        0 & \cos\theta_{23} & \sin\theta_{23}\, e^{-i \phi_{23}} \\
        0 & -\sin\theta_{23} \,e^{i \phi_{23}} & \cos\theta_{23}
        \end{matrix} \right) \,.
\end{equation}
The phases $\phi_{12}$ and $\phi_{13}$ are relevant for neutrinoless double beta decay while the combination $\delta_{CP} = \phi_{13} - \phi_{12} - \phi_{23}$ is the usual Dirac CP-violating phase measured in neutrino oscillations.

We now diagonalize the lepton mass matrices and show the existence of a neutrino mass sum rule.

\section{Neutrino mass sum rule and absolute neutrino mass scale}
\label{sec:sumrule}

\subsection{Derivation of the sum rule}

The physical neutrino masses correspond to the singular values of $\mathcal{M}_\nu$. Therefore, the basis-invariant traces of powers of $\mathcal{M}_\nu^\dagger \mathcal{M}_\nu$ directly encode the information of the mass spectrum. We compute the basis invariants related to the neutrino masses as \cite{Bento:2023owf}
\begin{eqnarray}
    I_{01} \equiv \text{Tr} \left(\mathcal{M}_\nu^\dagger \mathcal{M}_\nu\right) = \sum_i m_i^2 = 2 \alpha^2 u^2 .\\
    I_{02} \equiv \text{Tr} \left(\mathcal{M}_\nu^\dagger \mathcal{M}_\nu\mathcal{M}_\nu^\dagger \mathcal{M}_\nu\right) = \sum_i m_i^4 = \frac{I_{01}^2}{2} .  \label{eq:invariantsumrule}\\
   I_{03} \equiv \text{Tr}  \left(\mathcal{M}_\nu^\dagger \mathcal{M}_\nu\mathcal{M}_\nu^\dagger \mathcal{M}_\nu \mathcal{M}_\nu^\dagger \mathcal{M}_\nu\right) = \sum_i m_i^6 = \nonumber\\
 \frac{I_{01}^3}{4} \left(1 + 6 \cos^2\theta \, \sin^4\theta \, \cos^2\phi \, \sin^2\phi
 \right). \label{eq:I6}
\end{eqnarray}
As expected, the phases $\eta_i$ do not affect the neutrino masses, although they will play a role in the mixing angles and CP-violating phases discussed below. The Eq.~\eqref{eq:invariantsumrule} implies a linear relation among the physical neutrino masses once the ordering is determined, irrespective of the choice of model parameters, i.e.
\begin{eqnarray}
    \sum_i m_i^4 = \frac{1}{2} \left(\sum_i m_i^2\right)^2  \, \rightarrow \ m_3^2 = \left( m_2 \pm m_1 \right)^2 \,\\
 \implies   \text{(NO): } m_3 = m_1 + m_2, \quad
    \text{(IO): } m_2 = m_1 + m_3.
\end{eqnarray}
Alternatively, irrespective of the ordering,
\begin{align}
    m_{\rm heaviest}= \frac{1}{2} \sum_i m_i \ . 
    \label{eq:sumruleh}
\end{align}

We have also verified that the neutrino mass sum rule is preserved under one-loop RGE evolution in our setup, which suggests a symmetry protection. The first invariant $I_{01}$ is completely determined by the measured mass-squared differences ($\Delta m_{ij}^2$) and the sum rule of Eq.~\eqref{eq:sumruleh}. Indeed, numerically, allowing a $3\sigma$ uncertainty range for the $\Delta m_{ij}^2$ \cite{deSalas:2020pgw},

\begin{eqnarray}
    I_{01}^\text{NO} = \left(5.03 \pm 0.16\right) \, 10^{-3}  \text{ eV}^2, \quad
    I_ {01}^\text{IO} = \left(4.98 \pm 0.16\right) \, 10^{-3}  \text{ eV}^2.
\end{eqnarray}
And similarly for the invariant $I_{03}$
\begin{eqnarray}
    I_{03}^\text{NO} = \left(3.88 \pm 0.09\right) \, 10^{-8}  \text{ eV}^6, \quad I_ {03}^\text{IO} = \left(3.08 \pm 0.07\right) \, 10^{-8}  \text{ eV}^6.
\end{eqnarray}
Since $I_{01}$ and $I_{03}$ are determined for each ordering by the sum rule and the mass-squared differences, Eq.~\eqref{eq:I6} gives us a restriction on $\theta$ and $\phi$, leaving us with a single degree of freedom in the vev alignment of $\Delta$, which we parametrize as $t$. Indeed,
\begin{align}
    \frac{4 I_{03}}{I_{01}^3}-1 = 6 \cos \theta^2 \sin \theta^4 \cos\phi^2 \sin \phi^2 \rightarrow
    \left(\theta(t), \phi(t) \right), \, t\in[0, 1[ \ .\label{eq:implicit}
\end{align}
At this stage, the neutrino mass spectrum is completely fixed and the alignment angles $(\theta, \phi)$ are restricted to a one-dimensional manifold, parametrized by $t$. No information from lepton mixing has been used so far.

Let us do a parameter counting up to this point. In the charged lepton sector, we have 3 free parameters, \( \alpha_i \), which are fixed by the 3 charged lepton masses, up to a discrete choice of permutations. In the neutrino sector, there are $7$ parameters ($5$ relevant for neutrino masses and mixings): a global factor \( \alpha \, u \), a global phase \( \eta_1 \) that does not affect either the neutrino masses or mixing, 2 $\Delta$ alignment angles \( \theta \) and \( \phi \), and 2 additional phases \( \eta_2 \) and \( \eta_3 \). Fixing the invariant \( I_{01} \) through the measured mass squared differences $\Delta m_{21}^2$ and $\Delta m_{31}^2$ determines the combination \( \alpha u \), while \( I_{02} \) automatically fixes the sum rule. Solving for \( I_{06} \) allows us to express \( \theta \) and \( \phi \) in terms of a single free parameter \( t \), thus yielding the correct neutrino mass differences through Eq.~\eqref{eq:implicit}.
Finally, the 3 physical mixing angles and the 3 CP-violating phases have to be computed from the remaining 3 parameters: \( t \), \( \eta_1 \), and \( \eta_2 \). Therefore, this is indeed a highly predictive flavour model, which we will discuss in Sec.~\ref{sec:mixing}. We now briefly summarize the possible tests of the neutrino sum rule.

\subsection{Phenomenological tests of the sum rule and its predictions}

We now turn to a detailed discussion of neutrino mass sum rules and their phenomenological aspects.
Once Eq.~\eqref{eq:sumruleh} combined with the measured neutrino mass-squared differences \cite{deSalas:2020pgw}, the sum rule fully determines the absolute neutrino mass spectrum. 
\begin{align}
    \textbf{NO:} \nonumber \\
    & m_3 = m_1 + m_2, \nonumber \\
    & \Delta m_{21}^2 = 7.5 \times 10^{-5} \, \text{eV}^2, \quad \Delta m_{31}^2 = 2.55 \times 10^{-3} \, \text{eV}^2,  \nonumber \\
    &  m_1 = 0.0282 \, \text{eV}, \quad m_2 = 0.0295 \, \text{eV}, \quad m_3 = 0.0578 \, \text{eV}.   \\
    \textbf{IO:}  \nonumber \\
    & m_2 = m_1 + m_3,  \nonumber \\
    & \Delta m_{21}^2 = 7.5 \times 10^{-5} \, \text{eV}^2, \quad \Delta m_{31}^2 = -2.45 \times 10^{-3} \, \text{eV}^2,  \nonumber \\
    & m_3 = 7.5 \times 10^{-4} \, \text{eV}, \quad m_1 = 0.049 \, \text{eV}, \quad m_2 = 0.050 \, \text{eV}.
\end{align}

This result carries important implications for several ongoing and future experiments. Cosmological observations probe the sum of neutrino masses, and it is predicted very precisely by this sum rule. Considering the mass-squared differences within their $3\sigma$ ranges, the sum of neutrino masses is given as follows:
\begin{subequations}
\label{eq:summ_nu}
\begin{align}
\sum_i m_i^\text{NO}  \in [0.1138, 0.1176] \text{ eV} \label{eq:summno}, \\ 
\sum_i m_i^\text{IO}  \in [0.1007, 0.1041] \text{ eV} \label{eq:summio}.
\end{align}
\end{subequations}

These values are compatible with the Planck 2018 results \cite{Planck:2018vyg} as well as the latest results from the SPT-3G \cite{SPT-3G:2025bzu}. Its successor, the Euclid mission, is now operational and is expected to significantly improve the sensitivity to the sum of neutrino masses \cite{Amendola:2016saw}. However, cosmological constraints on the sum of neutrino masses are dependent on the considered data-sets and cosmological models \cite{Bertolez-Martinez:2024wez,Naredo-Tuero:2024sgf}.

Complementary constraints arise from direct neutrino mass measurements. Currently, two major experiments pursuing this approach are KATRIN \cite{KATRIN:2024cdt} and Project-8 \cite{Project8:2022hun,Project8:2025aar}. The effective electron neutrino mass probed in these experiments is defined as
\begin{align}
    m_{\rm eff}^{\nu_e} &= \sqrt{\sum_i |U_{ei}|^2\,m_i^2} \ .
\end{align}
KATRIN directly measures the effective mass of the electron neutrino and has recently reported the bound \cite{KATRIN:2024cdt}
\begin{align}
m_{\rm eff}^{\nu_e} &< 0.45~\text{eV},\quad 90\%\ \text{CL}~(\text{2024}).
\end{align}
In contrast, the current sensitivity of Project-8 is considerably weaker than that of KATRIN, yielding the limit \cite{Project8:2022hun}
\begin{align}
m_{\rm eff}^{\nu_e} < 155~\text{eV} \quad (\text{2022}).
\end{align}
However, future larger-volume Project-8 setups aim to reach sensitivities competitive with KATRIN, with projected reach down to a goal of $m_{\rm eff}^{\nu_e} < 0.04$ eV  \cite{Project8:2022wqh}.
If the mixing parameters are taken to their best fit values, the sum rule predicts
\begin{align}
\text{NO:}\quad m_{\rm eff}^{\nu_e} &= 0.028~\text{eV},\\
\text{IO:}\quad m_{\rm eff}^{\nu_e} &= 0.049~\text{eV}.
\end{align}
Therefore, if KATRIN or Project-8 measure a nonzero neutrino mass above these values during their runs, the model would be ruled out. Additionally, the future large volume version of Project-8 could completely probe the sum rule prediction for IO.
Furthermore, this sum rule also has important consequence for the $0 \nu e e$ beta decay as we discuss in Sec.~\ref{sec:mixing}.

Up to this point we have obtained a series of powerful analytical results that can be extracted from the model. In what follows we perform a numerical scan over the remaining $3$ free parameters $t$, $\eta_2$ and $\eta_3$, as well as the discrete permutation choices in the charged lepton sector and neutrino mass ordering.

\section{Predictions for neutrino mass ordering, Lepton mixing, CP Violation and $\mathbf{0 \nu e e}$}
\label{sec:mixing}
\subsection{Normal ordering}

The mixing angles predictions for the normal ordering of neutrino masses are inconsistent with the neutrino global fit \cite{deSalas:2020pgw}. Therefore, the model predicts IO of the neutrino mass. The already ongoing experiment JUNO \cite{Khan:2019doq, JUNO:2025gmd} will be able to determine the ordering of neutrino masses and thus probe this model prediction. We now turn our attention to the IO.

\subsection{Inverted Ordering}

\begin{figure}[!h]
  \centering
  \includegraphics[width=0.55\textwidth]{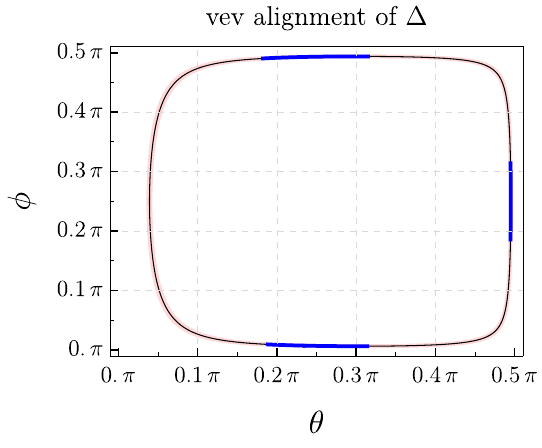}
\caption{Region in the $(\theta,\phi)$ plane satisfying the neutrino mass constraints. $(\theta,\phi)$ are the $\Delta$ vev alignment angles, see Eq.~\eqref{eq:alignment}. The black contour corresponds to the best-fit values of the neutrino mass-squared differences~\cite{deSalas:2020pgw}, while the pink band shows the corresponding $3\sigma$ allowed range. The blue segments, artificially thickened for visibility, indicate the subset of solutions that are also compatible with the observed lepton mixing angles. All viable solutions lie close to the limiting vev alignments $(1,1,0)$, $(1,0,1)$, and $(0,1,1)$.}
  \label{fig:alignment}
\end{figure}

Continuing from Eq.~\eqref{eq:implicit}, and allowing the mass-squared differences to vary within their $3\sigma$ ranges, we have
\begin{equation}
    \left(\frac{4 I_{03}}{I_{01}^3}-1\right)^\text{IO} = \left(3.4 \pm 0.6\right)\, \times 10^{-4} .
\end{equation}
Numerically scanning over $t$, $\eta_1$ and $\eta_2$ and fitting the lepton mixing angles on their $3\sigma$ allowed ranges, we find that the vev alignment must be a perturbation of the vaccum $(1,1,0)$ (and permutations), see Fig.~\ref{fig:alignment}. 
This result is purely phenomenological and follows from the interplay between the mass sum rule and the mixing constraints. The dynamical realization of such alignments is discussed in Appendix \ref{app:scalarpot}.

\begin{figure} [!h]
    \centering
    \includegraphics[width=0.47\linewidth]{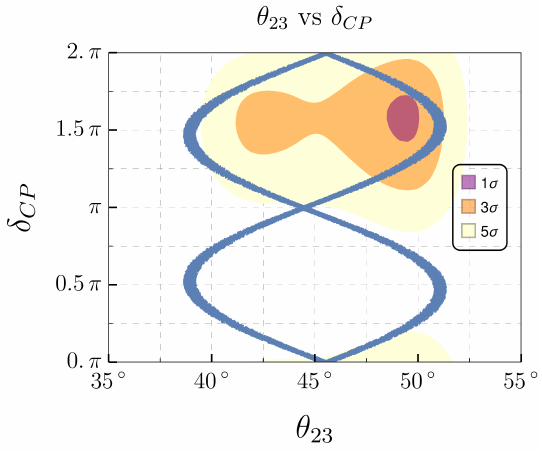}
    \includegraphics[width=0.47\linewidth]{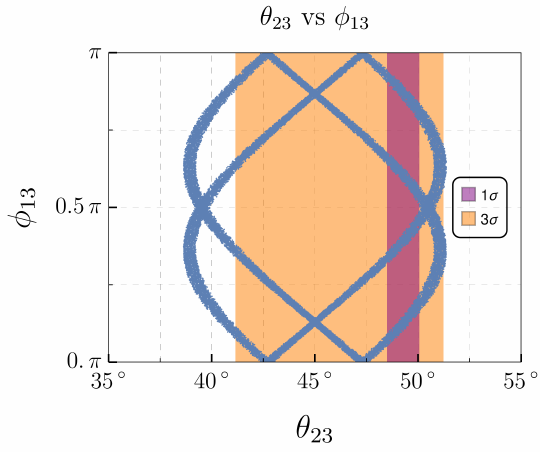}
    \caption{Correlation of the atmospheric mixing angle $\theta_{23}$, left panel: with the Dirac CP-violating phase $\delta_{\rm CP}$ and right panel: Majorana phase $\phi_{13}$.}
    \label{fig:delcp}
\end{figure}

The model also features a non-trivial correlation among oscillation parameters. Imposing the current $3\sigma$ allowed ranges for $\theta_{13}$ and $\theta_{12}$, while leaving $\theta_{23}$ and the Dirac CP-violating phase $\delta_{CP}$ free, we find a strong correlation between $\theta_{23}$ and $\delta_{CP}$, shown in the left panel of Fig.~\ref{fig:delcp}. Remarkably, the model predicts $\theta_{23}$ to lie close to the maximal mixing value, in good agreement with current neutrino oscillation data.
Interestingly, these two parameters are currently the least precisely measured among the oscillation observables. In particular, present data show a mild tension in the preferred values of $\delta_{CP}$ as extracted by NOvA \cite{NOvA:2021nfi} and T2K \cite{T2K:2023smv}.
The correlation predicted by the model therefore provides a clear target for future measurements.
Significant progress in the determination of both $\theta_{23}$ and $\delta_{CP}$ is expected from next-generation long-baseline experiments, most notably DUNE \cite{DUNE:2020ypp} and Hyper-Kamiokande \cite{Hyper-Kamiokande:2018ofw}, which are designed to substantially reduce the uncertainties on these parameters. As a result, the $\theta_{23}$–$\delta_{CP}$ correlation predicted here will be decisively tested in the near future.

Furthermore, there is a correlation between the CP-Violating phases of the model, two of which are not experimentally determined (Majorana phases, $\phi_{12}$ and $\phi_{13}$), and the other mixing parameters. As an example, see $\theta_{23}$ vs $\phi_{13}$ in Fig.~\ref{fig:delcp} right panel.
\begin{figure} [!h]
    \centering
    \includegraphics[width=0.47\linewidth]{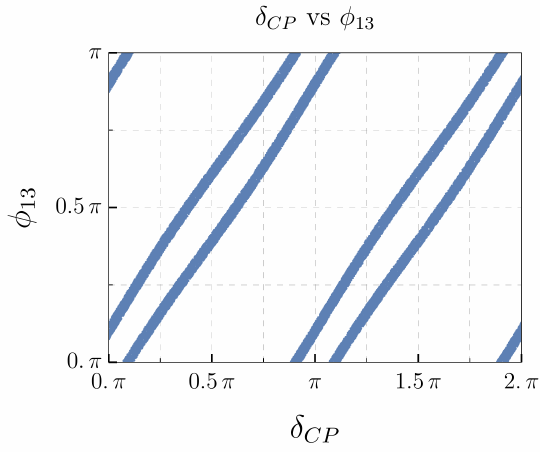}
    \includegraphics[width=0.47\linewidth]{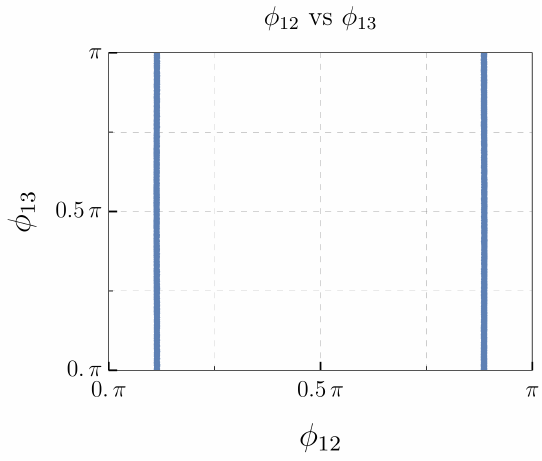}
    \caption{Correlations between CP-violating phases. Left panel: $\delta_{CP}$ vs $\phi_{13}$. Right panel: $\phi_{12}$ vs $\phi_{13}$. Note that the Majorana phase $\phi_{12}$ is forced to be close to the values maximizing $|m_{ee}|$, see Eqs.~\eqref{eq:phi12} and \eqref{eq:mee}.}
    \label{fig:phases}
\end{figure}
Additionally, $\phi_{12}$ is approximately restricted to $2$ fixed values, see Fig.~\ref{fig:phases},
\begin{equation}
    \phi_{12} \approx 0.11 \, \pi \text{ or } 0.89\,\pi . \label{eq:phi12}
\end{equation}

\begin{figure} [!h]
    \centering
\includegraphics[width=0.5\linewidth]{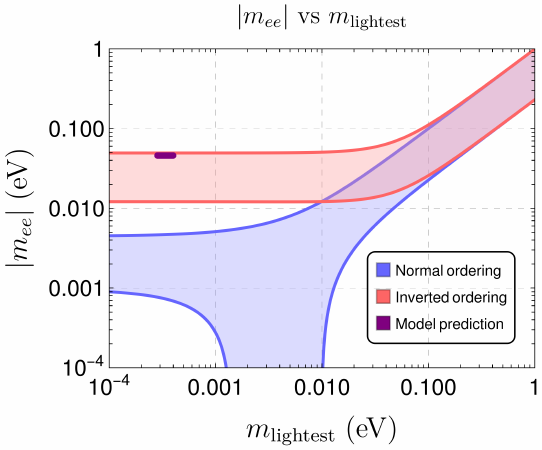}
    \caption{Effective Majorana mass $|m_{ee}|$ as a function of the lightest neutrino mass $m_{\text{lightest}}$. The blue and red bands correspond to the allowed regions for normal and inverted mass ordering, respectively, obtained by varying the oscillation parameters within their current $3\sigma$ ranges. The purple band shows the prediction of the model, which enforces inverted ordering sum rule and restricts the Majorana phase $\phi_{12}$ (see Eq.~\eqref{eq:phi12}), leading to a highly localized region for $|m_{ee}|$. This results in a sharp and testable prediction for neutrinoless double beta decay experiments.}
    \label{fig:mee}
\end{figure}

This has consequences for neutrinoless beta decay experiments, since the rate of this process is proportional to the quantity $|m_{ee}|$, given by

\begin{equation}
 \label{eq:mee}
|m_{ee}| = \left|\sum_i U_{ei}^2 m_i\right| = |c_{12}^2 c_{13}^2 m_1 + s_{12}^2 c_{13}^2 e^{2 i \phi_{12}} m_ 2 + s_{13}^2 e^{2 i \phi_{13}} m_3| \, .
\end{equation}
For $m_3 \ll m_2, m_1$, which is the case for IO with the sum rule, Eq.~\eqref{eq:mee} is maximized for $\phi_{12} = 0, \pi$. The model predicts Eq.~\eqref{eq:phi12}, with values close to  these extrema, leading to a sharply localized prediction for $|m_{ee}|$, see Fig.~\ref{fig:mee}. 
In particular, the sharp prediction of the model is
\begin{eqnarray}
    |m_{ee}|^\text{prediction} \approx 46 \text{ meV}.
\end{eqnarray}
To be compared with the latest KamLAND-Zen result \cite{KamLAND-Zen:2024eml}, which at 90\% CL level restricts $|m_{ee}| < (28-122) \ \text{meV}$, depending on the uncertainties of the nuclear matrix elements. The successor experiment, KamLAND-Zen2 \cite{Nakane:2025yxq} may completely rule out or confirm the model prediction.

\section{Charged Lepton Flavour Violation} \label{sec:LFV}
Charged lepton flavour-violating processes offer stringent constraints on new physics scenarios, with current experimental limits placing strong bounds on radiative and three-body decays \cite{ParticleDataGroup:2024cfk}. In our model, cLFV processes are mediated by both the $H$ and $\Delta$ flavour triplets, whose interactions generally induce flavour violation in the charged lepton sector.
However, a particularly simple structure emerges in the \textit{seesaw limit}, in which the components of $\Delta$ become heavy compared to the electroweak scale and effectively decouple. In this limit, the low-energy theory recovers a triality symmetry~\cite{Ma:2001dn,Ma:2010gs,Cao:2011df} that governs the remaining Higgs-mediated interactions and sharply constrains the flavour structure of cLFV processes. Throughout this section, we work in this heavy-$\Delta$ regime, where the dominant LFV effects can be analyzed in a controlled and transparent manner.
To analyze the possible LFV decay rates, we first rotate the charged lepton and Higgs interaction terms given in Eq.~\eqref{eq:lag2} from the interaction basis to the physical mass basis. In the limit $u \ll v$, which is equivalent to $m_\nu \ll \Lambda_{EW}$, the mass matrices of $H$ and $\Delta$ do not mix. In the mass basis, Eq.~\eqref{eq:lag2} can be written as follows.
\begin{align} \label{eq:lagmass}
    \mathcal{L}_{int}&= \frac{1}{v} \left[m_{\tau} \bar{L}_{\tau} \tau_{R} + m_{\mu} \bar{L}_{\mu} \mu_{R} + m_{e} \bar{L}_{e} e_{R}  \right] \phi_0 \nonumber \\ 
    & + \frac{1}{v} \left[m_{\tau} \bar{L}_{\mu} \tau_{R} + m_{\mu} \bar{L}_{e} \mu_{R} + m_{e} \bar{L}_{\tau} e_{R}  \right] \phi_1 \nonumber \\
    & + \frac{1}{v} \left[m_{\tau} \bar{L}_{e} \tau_{R} + m_{\mu} \bar{L}_{\tau} \mu_{R} + m_{e} \bar{L}_{\mu} e_{R}  \right] \phi_2  + \text{h.c.} ,
\end{align}
where $\phi_0$,  $\phi_1$, and $\phi_2$ are the mass eigenstates and are related with $H_1$, $H_2$, and $H_3$ as
\begin{align}
    \begin{pmatrix}
        \phi_0 \\
        \phi_1 \\
        \phi_2
    \end{pmatrix}= \frac{1}{\sqrt{3}} \begin{pmatrix}
        1 & 1 &1 \\
        1 & \omega & \omega^2 \\
         1 & \omega^2 & \omega
    \end{pmatrix}  \begin{pmatrix}
        H_1 \\
        H_2 \\
        H_3
    \end{pmatrix}.
\end{align}

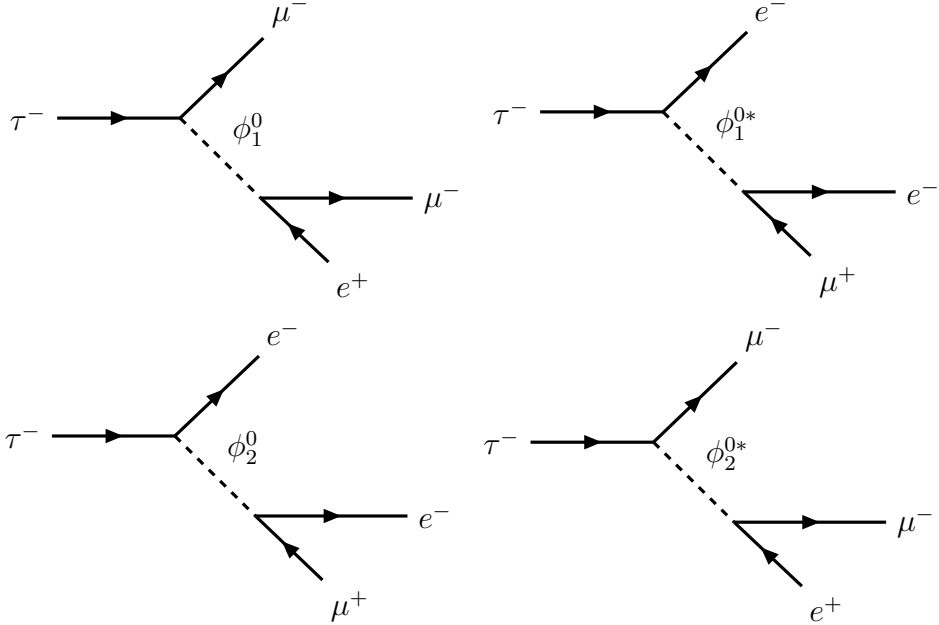
\begin{figure}[!h]
    \centering
    \begin{tikzpicture}
\begin{feynman}

\vertex (mu){\(\tau^-\)};
\vertex[right=2.0cm of mu] (v1);
\vertex[above right=1.5cm of v1] (nu_mu){\(\mu^-\)};
\vertex[below right=1.5cm of v1] (v2);
\vertex[right=2.0cm of v2] (e) {\(\mu^-\)};
\vertex[below right=1.2cm of v2] (nu_e){\(e^+\)};

\diagram*{

(mu) -- [fermion, very thick] (v1),

(v1) -- [scalar, very thick, edge label=\(\phi_1^0\)] (v2),

(v1) -- [fermion, very thick] (nu_mu),
(v2) -- [fermion, very thick] (e),
(nu_e) -- [fermion, very thick] (v2),

};

\end{feynman}
\end{tikzpicture}
\begin{tikzpicture}
\begin{feynman}

\vertex (mu){\(\tau^-\)};
\vertex[right=2.0cm of mu] (v1);
\vertex[above right=1.5cm of v1] (nu_mu){\(e^-\)};
\vertex[below right=1.5cm of v1] (v2);
\vertex[right=2.0cm of v2] (e) {\(e^-\)};
\vertex[below right=1.2cm of v2] (nu_e){\(\mu^+\)};

\diagram*{

(mu) -- [fermion, very thick] (v1),

(v1) -- [scalar, very thick, edge label=\(\phi_1^{0 \ast}\)] (v2),

(v1) -- [fermion, very thick] (nu_mu),
(v2) -- [fermion, very thick] (e),
(nu_e) -- [fermion, very thick] (v2)  ,

};

\end{feynman}
\end{tikzpicture}
\begin{tikzpicture}
\begin{feynman}

\vertex (mu){\(\tau^-\)};
\vertex[right=2.0cm of mu] (v1);
\vertex[above right=1.5cm of v1] (nu_mu){\(e^-\)};
\vertex[below right=1.5cm of v1] (v2);
\vertex[right=2.0cm of v2] (e) {\(e^-\)};
\vertex[below right=1.2cm of v2] (nu_e){\(\mu^+\)};

\diagram*{

(mu) -- [fermion, very thick] (v1),

(v1) -- [scalar, very thick, edge label=\(\phi_2^0\)] (v2),

(v1) -- [fermion, very thick] (nu_mu),
(v2) -- [fermion, very thick] (e),
(nu_e) -- [fermion, very thick] (v2),

};

\end{feynman}
\end{tikzpicture}
\begin{tikzpicture}
\begin{feynman}

\vertex (mu){\(\tau^-\)};
\vertex[right=2.0cm of mu] (v1);
\vertex[above right=1.5cm of v1] (nu_mu){\(\mu^-\)};
\vertex[below right=1.5cm of v1] (v2);
\vertex[right=2.0cm of v2] (e) {\(\mu^-\)};
\vertex[below right=1.2cm of v2] (nu_e){\(e^+\)};

\diagram*{

(mu) -- [fermion, very thick] (v1),

(v1) -- [scalar, very thick, edge label=\(\phi_2^{0 \ast}\)] (v2),

(v1) -- [fermion, very thick] (nu_mu),
(v2) -- [fermion, very thick] (e),
(nu_e) -- [fermion, very thick] (v2) ,

};

\end{feynman}
\end{tikzpicture}
    \caption{Schematic Feynman diagrams of the allowed $\tau^-$ three-body decay channels mediated by the scalars $\phi^0_1$ and $\phi^0_2$.}
    \label{fig:LFV}
\end{figure}

The interaction term involving $\phi_0$ is flavour diagonal and hence does not induce LFV processes. On the other hand, the couplings associated with $\phi_1$ and $\phi_2$ generate flavour off-diagonal interactions and are therefore responsible for LFV decay processes. Remarkably, all radiative two-body LFV decays, including $\mu \to e \gamma$, $\tau \to \mu \gamma$, and $\tau \to e \gamma$, are absent in this framework. As a consequence, the leading LFV signatures arise from three-body decays of the form $l_i^- \to l_j^- l_k^- l_m^+$. Among these, only the following processes are allowed:
\begin{align}
    & \ \mathcal{P}_1: \tau^- \to \mu^- \mu ^- e^+ , \quad \mathcal{P}_2: \tau^- \to  e^- e^- \mu^+ \   \quad \text{(mediated by $\phi_1^0$)} \nonumber \\
     & \ \mathcal{P}_3: \tau^- \to e^- e^- \mu^+ \ , \quad \mathcal{P}_4: \tau^- \to \mu^- \mu ^- e^+ \  \quad \text{(mediated by $\phi_2^0$)} 
\end{align}
where $\phi_{1,2}^0$ are the neutral part of the scalars $\phi_{1,2}$. The Feynman diagrams for these channels have been shown in Fig.~\ref{fig:LFV}.
The associated decay widths are proportional to $m_\tau m_\mu / v^2$ and $m_e m_\mu / v^2$ for $\phi^0_1$ mediation, while for $\phi^0_2$ mediation they scale as $m_\tau m_e / v^2$ and $m_e m_\mu / v^2$, respectively. 
The decay width $(\Gamma_{\mathcal{P}_i})$ for these channels can be computed as follows: 
\begin{align}
    &\Gamma_{\mathcal{P}_1}= \frac{1}{192 \pi^3}\frac{m_{\tau}^5}{m_{\phi^0_1}^4} \left( \frac{m_{\tau} m_{\mu}}{v^2}\right)^2, \quad \Gamma_{\mathcal{P}_2}= \frac{1}{192 \pi^3}\frac{m_{\tau}^5}{m_{\phi^0_1}^4} \left( \frac{m_{e} m_{\mu}}{v^2}\right)^2, \nonumber \\
    &\Gamma_{\mathcal{P}_3}= \frac{1}{192 \pi^3}\frac{m_{\tau}^5}{m_{\phi^0_2}^4} \left( \frac{m_{\tau} m_{e}}{v^2}\right)^2, \quad \Gamma_{\mathcal{P}_4}= \frac{1}{192 \pi^3}\frac{m_{\tau}^5}{m_{\phi^0_2}^4} \left( \frac{m_{e} m_{\mu}}{v^2}\right)^2 \ .
\end{align}

Since, the $\Gamma (\tau^- \to \mu^- \overline{\nu}_{\mu} \nu_{\tau}) \approx \frac{G^2_F}{192 \pi^3} m_{\tau}^5$ ($G_F$ is the Fermi constant), the branching ratios $(\mathcal{B R})$ of these processes ($\mathcal{P}_i$) are given as follows:
\begin{align}
    &\mathcal{B R} (\mathcal{P}_1) = \left( \frac{m_{\tau} m_{\mu}}{v^2}\right)^2 \left(\frac{1}{G^2_F m_{\phi^0_1}^4} \right) \mathcal{B R} (\tau^- \to \mu^- \overline{\nu}_{\mu} \nu_{\tau}), \nonumber \\ 
    &\mathcal{B R} (\mathcal{P}_2) = \left( \frac{m_{e} m_{\mu}}{v^2}\right)^2 \left(\frac{1}{G^2_F m_{\phi^0_1}^4} \right) \mathcal{B R} (\tau^- \to \mu^- \overline{\nu}_{\mu} \nu_{\tau}), \nonumber \\ 
    &\mathcal{B R} (\mathcal{P}_3) = \left( \frac{m_{\tau} m_{e}}{v^2}\right)^2 \left(\frac{1}{G^2_F m_{\phi^0_2}^4} \right) \mathcal{B R} (\tau^- \to \mu^- \overline{\nu}_{\mu} \nu_{\tau}), \nonumber \\ 
    &\mathcal{B R} (\mathcal{P}_4) = \left( \frac{m_{e} m_{\mu}}{v^2}\right)^2 \left(\frac{1}{G^2_F m_{\phi^0_2}^4} \right) \mathcal{B R} (\tau^- \to \mu^- \overline{\nu}_{\mu} \nu_{\tau}).
\end{align}
Except for $\mathcal{P}_1$, all other processes are suppressed by the electron mass $m_e^2$. Moreover, due to the triality symmetry, the three-body decay $\mu \to e e e$ is also forbidden. The branching ratios shown above have a lower bound within the model. We focus on the dominant channel $\mathcal{P}_1$. Since the electroweak scale receives contributions from multiple scalar multiplets, the vev $v$ associated with the flavour triplet $H$ satisfies $v<v_{SM}$. Moreover, assuming perturbative $\lambda$ couplings in the scalar potential we have $m_{\phi_1^0} \lsim v < v_{SM}$, implying
\begin{small}
\begin{equation}
    \mathcal{B R} (\mathcal{P}_1) = \left( \frac{m_{\tau} m_{\mu}}{v^2}\right)^2 \left(\frac{1}{G^2_F m_{\phi^0_1}^4} \right) \mathcal{B R} (\tau^- \to \mu^- \overline{\nu}_{\mu} \nu_{\tau}) > \left( \frac{m_{\tau} m_{\mu}}{v_{SM}^2}\right)^2 \left(\frac{1}{G^2_F v_{SM}^4} \right) \mathcal{B R} (\tau^- \to \mu^- \overline{\nu}_{\mu} \nu_{\tau})
\end{equation}
\end{small}
Using the current PDG \cite{ParticleDataGroup:2024cfk} values for the charged lepton masses and electroweak parameters, we obtain the lower bound for the branching ratio of the process $\mathcal{P}_1$

\begin{eqnarray}
    \mathcal{B R} (\mathcal{P}_1) \gsim  10^{-12}.
\end{eqnarray}
While the experimental limit, also taken from the PDG, is $\mathcal{B R} (\mathcal{P}_1) < 1.7 \times 10^{-8}$. cLFV is therefore not expected to provide the leading experimental test of the model. On the other hand, triality-breaking muon decays such as $\mu \to e \gamma$ are experimentally better measured, and their detection, understood in the context of the current model, would signal a relatively low scale of the triplet $\Delta$.

\section{Conclusions}
\label{sec:conclusions}
In this work, we have presented a highly predictive flavour model based on an $A_4$ symmetry combined with a type-II seesaw mechanism for neutrino mass generation. The flavour structure of neutrino mass matrix leads to a neutrino mass sum rule written directly in terms of the \emph{physical} neutrino masses, understood as the singular values of the neutrino mass matrix. Once the measured mass-squared differences are imposed, this sum rule completely fixes the absolute neutrino mass spectrum, independently of model parameters. 

After imposing current neutrino oscillation constraints, the model also predicts \emph{inverted neutrino mass ordering}. This feature will be robustly tested by upcoming medium-baseline reactor experiments such as JUNO, which are expected to determine the neutrino mass ordering with high significance. In addition, the sum rule leads to sharply predicted values for the effective electron neutrino mass $m_{\nu_e}^{\rm eff}$, placing the model squarely within the reach of future direct kinematic searches. In particular, the ultimate sensitivity goals of the Project~8 experiment is expected to probe the parameter space relevant for the inverted ordering sum rule predicted here, providing a direct and model-independent test of the sum rule.

The constrained flavour structure also leads to non-trivial predictions for lepton mixing. While the solar and reactor mixing angles are imposed within their current $3\sigma$ ranges, the atmospheric mixing angle $\theta_{23}$ and the Dirac CP-violating phase $\delta_{CP}$ emerge as correlated predictions of the model. Remarkably, $\theta_{23}$ is predicted to lie close to maximal mixing, in good agreement with present data. Since $\theta_{23}$ and $\delta_{CP}$ are currently the least precisely measured oscillation parameters, the predicted correlation provides a clear and testable target for next-generation long-baseline experiments, such as DUNE and Hyper-Kamiokande.

The model further predicts a highly localized region for the effective Majorana mass $|m_{ee}|$ governing neutrinoless double beta decay. Due to the restricted values of the Majorana phases enforced by the flavour structure, the predicted
$|m_{ee}|$ lies close to its maximal value for inverted ordering. Interestingly, the current KamLAND-Zen limits already probe part of this region, although dominated by uncertainties in the nuclear matrix elements. The upcoming KamLAND-Zen2 experiment is therefore expected to play a decisive role in either confirming or ruling out this prediction.

Finally, charged lepton flavour violation provides a complementary probe of the model. In the seesaw limit, where the scalar triplet $\Delta$ is heavy and decoupled, the low-energy theory exhibits lepton triality, leading to the absence of muon flavour-violating decays, while allowing specific three-body $\tau$ decays. Although the predicted $\tau$ decay rates lie well below current experimental sensitivities, the observation of triality-breaking processes such as $\mu \to e\gamma$ or $\mu \to eee$ would provide indirect evidence for a relatively low scale of the triplet $\Delta$, offering an alternative window into the ultraviolet completion of the model.

The model yields a set of predictions for neutrino mass ordering, absolute neutrino mass measurements, CP violation, neutrinoless double beta decay and charged lepton flavour violation. The convergence of multiple experimental frontiers in the coming years therefore makes this framework both highly constrained and testable. 

\section*{Acknowledgments}

SCCh acknowledges support from the Spanish grants PID2023-147306NB-I00, CNS2024-154524 and CEX2023-001292-S (MICIU/AEI/10.13039/501100011033).
RK would like to acknowledge support from the Research Program funded by the Seoul National University of Science and Technology.
We would also like to thank the organizers of the FLASY 2025 conference in Rome, where this project was initiated.

\appendix

\section{Quark Mass Generation} \label{app:quarkmass}
In this section, we briefly address the generation of quark masses within our framework. An additional Higgs-like doublet $H_q$, singlet under $A_4$, is introduced for this purpose. All quark fields are likewise assigned as $A_4$ singlets, ensuring that their Yukawa interactions involve only $H_q$. This construction preserves the flavour structure of the lepton sector while providing a straightforward and consistent origin for quark masses. The relevant Yukawa Lagrangian for the quark sector can be formulated as
\begin{align} \label{eq:quarklag}
    -\mathcal{L}_q = y^u_{ij} \bar{Q} \tilde{H}_q u_j + y^d_{ij} \bar{Q} H_q d_j + \rm{h.c.}, 
\end{align}
where $Q$ denotes the left handed quark doublets, while $u$ and $d$ represent the right handed up and down type quark singlets under $SU(2)_L$, respectively, and $\tilde{H}_q \equiv i \tau_2 H_q^{\ast}$. The Yukawa structure encoded in Eq.~\eqref{eq:quarklag} is sufficient to account for the observed quark masses and mixings, after the scalar $H_q$ acquires vev, $\langle H_q \rangle = v_q $.

\section{Scalar Sector and Mass Spectrum} \label{app:scalarpot}
The scalar sector of the model contains a $SU(2)_L$ doublet Higgs $H$ and a triplet $\Delta$, both of which are triplets under the $A_4$ flavour symmetry. In addition, a separate Higgs doublet $H_q$, singlet under $A_4$, is introduced to generate the
quark masses.
Since $H_q$ is a flavour singlet and appears only in the quark sector, it does not participate in the flavour dynamics of the lepton sector discussed in this work. Therefore, for clarity, we present a separate discussion of the scalar potential involving $H_q$.
The scalar potential pertaining to $H$ and $\Delta$, invariant under the SM gauge group and $A_4$,  can be formulated as follows~\cite{Buskin:2021eig}:
\begin{align}
\label{eq:pot3}
V&=-m_{H}^2 \left(H_1^\dagger H_1 + H_2^\dagger H_2 + H_3^\dagger H_3 \right) 
+ \lambda_{H1} \left(H_1^\dagger H_1 + H_2^\dagger H_2 + H_3^\dagger H_3 \right)^2
\nonumber \\ &+ \lambda_{H2} \left[\left(H_1^\dagger H_1 \right)\left(H_2^\dagger H_2\right) + \left(H_2^\dagger H_2\right)\left(H_3^\dagger H_3\right) + \left(H_3^\dagger H_3\right)\left(H_1^\dagger H_1\right) \right] \nonumber \\ &+ \lambda_{H3} \left(|H_1^\dagger H_2|^2 + |H_2^\dagger H_3|^2 + |H_3^\dagger H_1|^2 \right) 
+ \lambda_{H4} \left[\left(H_1^\dagger H_2  \right)^2 + \left(H_2^\dagger H_3  \right)^2 + \left(H_3^\dagger H_1  \right)^2 + \text{h.c.} \right]
\nonumber \\ 
&- M_{\Delta}^{'2} \text{Tr} \left[\left(\Delta_1^\dagger\Delta_1 + \Delta_2^\dagger\Delta_2 + \Delta_3^\dagger\Delta_3 \right) \right] +
\lambda_{\Delta 0} \left(\text{Tr} \left[\Delta_1^\dagger\Delta_1 + \Delta_2^\dagger\Delta_2 + \Delta_3^\dagger\Delta_3 \right]\right)^2 \nonumber \\&
+ \lambda_{\Delta 1} \text{Tr} \left[\left(\Delta_1^\dagger\Delta_1 + \Delta_2^\dagger\Delta_2 + \Delta_3^\dagger\Delta_3\right)^2 \right] \nonumber \\&
+  \lambda_{\Delta 2} \text{Tr} \left[\left(\Delta_1^\dagger\Delta_1 \right)\left(\Delta_2^\dagger\Delta_2 \right)+\left(\Delta_2^\dagger\Delta_2 \right)\left(\Delta_3^\dagger\Delta_3 \right)+\left(\Delta_3^\dagger\Delta_3 \right)\left(\Delta_1^\dagger\Delta_1 \right) \right] 
\nonumber  \\&+
 \lambda_{\Delta 3} \text{Tr} \left[|\Delta_1^\dagger\Delta_2|^2+ |\Delta_2^\dagger\Delta_3|^2 + |\Delta_3^\dagger\Delta_1|^2 \right]
+  \lambda_{\Delta 4} \text{Tr}\left[ \left(\Delta_1^\dagger\Delta_2\right)^2 + \left(\Delta_2^\dagger\Delta_3 \right)^2 + \left(\Delta_3^\dagger\Delta_1 \right)^2 + \text{h.c.} \right ] 
\nonumber  \\&+\lambda_{H \Delta 1}\left(H_1^\dagger H_1 + H_2^\dagger H_2 + H_3^\dagger H_3 \right)\text{Tr}\left[\left(\Delta_1^\dagger\Delta_1 + \Delta_2^\dagger\Delta_2 + \Delta_3^\dagger\Delta_3 \right)\right]
\nonumber \\
&+\lambda_{H \Delta 2}\left[\left(H_1^\dagger \Delta_2\right)\left(\Delta_2^\dagger H_1\right) + \left(H_2^\dagger \Delta_3\right)\left(\Delta_3^\dagger H_2\right) +\left(H_3^\dagger \Delta_1\right)\left(\Delta_1^\dagger H_3\right)\right] 
\nonumber \\
&+\lambda_{H \Delta 3}\left[\left(H_1^\dagger \Delta_2\right)\left(\Delta_1^\dagger H_2\right) + \left(H_2^\dagger \Delta_3\right)\left(\Delta_2^\dagger H_3\right) +\left(H_3^\dagger \Delta_1\right)\left(\Delta_3^\dagger H_1\right) +\text{h.c.}\right] 
\nonumber \\
&+ \kappa_1 \left( H_1^T \tilde{\Delta}^\dagger_2 H_3 + H_2^T \tilde{\Delta}^\dagger_3 H_1 + H_3^T \tilde{\Delta}^\dagger_1 H_2 +\text{h.c.} \right) + \kappa_2 \left( H_1^T \tilde{\Delta}^\dagger_3 H_2 + H_2^T \tilde{\Delta}^\dagger_1 H_3 + H_3^T \tilde{\Delta}^\dagger_2 H_1 +\text{h.c.} \right)
  \;.
\end{align}
where $ \tilde{\Delta_i}\equiv i \tau_2 \Delta_i$. As discussed in Sec.~\ref{sec:mixing}, all viable solution for the vev alignment of $\Delta$ are close to $u_i=u_j=u'$, $u_k=0$. Thus the tadpole solutions with this approximation are given by
\begin{align}
\mu_H^2 =&
 (3\lambda_{H1}
+ \lambda_{H2}
+ \lambda_{H3}
+ 2\lambda_{H4})v^2
+ \frac{1}{2}(2 \lambda_{H\Delta1}
+ \lambda_{H\Delta2}
+ \lambda_{H\Delta3})u'^2
 -\frac{\kappa_1 + \kappa_2 }{\sqrt{2}} u',
\nonumber \\
M_{\Delta}^{'2} =& \
(2\lambda_{\Delta 0}
+2\lambda_{\Delta 1}+\frac{\lambda_{\Delta 2}}{2}
+\frac{\lambda_{\Delta 3}}{2}
+\lambda_{\Delta4})u'^2 +
\frac{1}{2}(3\lambda_{H\Delta1}
+\lambda_{H\Delta2}
+\lambda_{H\Delta3})v^2
-\frac{\kappa_1 + \kappa_2}{\sqrt{2}u'} v^2
. 
\end{align}
The exact alignment $(u,u,0)$ follows from the scalar potential in Eq.~\eqref{eq:pot3}. Small deviations from this configuration, required to reproduce the phenomenologically viable alignment discussed in Sec.~\ref{sec:mixing}, can be obtained by introducing soft $A_4$-breaking terms in the scalar potential of the form $\mu_{ij}\,\Delta_i^\dagger \Delta_j$.
These terms do not affect the structure of the Yukawa sector discussed in the main text and can naturally be taken small, ensuring that the alignment remains close to the symmetric configuration.

In the limit $u \ll v$, the  Higgs mass squared matrix effectively decouples from the $\Delta$ mass squared matrix and can be written as
\begin{align}
    \mathcal{M}_H^2= v^2\begin{pmatrix}
        a & a+b & a+b \\
        a+b & a &  a+b \\
         a+b &  a+b & a
    \end{pmatrix}.
\end{align}
The eigenvalues of this matrix are as follows
\begin{align}
    m^2_h = (3a+2b)v^2, \quad m^2_{h_2, h_3}=  -b v^2,  
\end{align}
where $a\equiv 2 \lambda_{H_1}$ and $b \equiv \lambda_{H_2} + \lambda_{H_3} + 2 \lambda_{H_4}$. 
The bounded from below (BFB) condition for the Higgs potential is given by \cite{Buskin:2021eig} 
\begin{align}
    &\lambda_{H_1} \geq 0, \quad 3 \lambda_{H_1} + \lambda_{H_2} + \lambda_{H_3} + 2 \lambda_{H_4} \geq 0, \nonumber\\
    &3 \lambda_{H_1} + \lambda_{H_2} + \lambda_{H_3} - \lambda_{H_4} \geq 0, \quad 4 \lambda_{H_1} + \lambda_{H_2} + \lambda_{H_3} - 2 \lambda_{H_4} \geq 0.
\end{align}

The inclusion of the additional scalar doublet $H_q$ modifies the scalar potential given in Eq.~\eqref{eq:pot3}. In the seesaw limit, the scalars $\Delta_i$ remain effectively decoupled. Consequently, we present only the additional relevant terms to the potential, which can be written as follows
\begin{align} \label{eq:potadd}
  V_{\rm add}= &- \mu^2_{H_q} H^{\dagger}_{q} H_q + \lambda_{H_{q}} \left ( H^{\dagger}_{q} H_q \right)^2 + \lambda'_2 \left(H^{\dagger}_{q} H_q \right) \left( H^\dagger_1 H_1 + H^\dagger_2 H_2+ H^\dagger_3 H_3 \right)
    \nonumber \\ &+ \lambda'_3 \left( |H^\dagger_q H_1|^2 + |H^\dagger_q H_2|^2 + |H^\dagger_q H_3|^2 \right) + \lambda'_4 \left[ (H^\dagger_q H_1)^2 + (H^\dagger_q H_2)^2 + (H^\dagger_q H_3)^2 + \rm{h.c.} \right]
     \nonumber \\ &+ \lambda'_5 \left[ (H^\dagger_q H_1)(H^\dagger_2 H_3) + (H^\dagger_q H_2)(H^\dagger_3 H_1) + (H^\dagger_q H_3)(H^\dagger_1 H_2) + \rm{h.c.} \right ] %
     \nonumber \\ & + \lambda'_6 \left[ (H^\dagger_q H_1)(H^\dagger_3 H_2) + (H^\dagger_q H_2)(H^\dagger_1 H_3) + (H^\dagger_q H_3)(H^\dagger_2 H_1) + \rm{h.c.} \right]
\end{align}

Exact SM-like behavior of the lightest CP-even scalar is obtained in the alignment limit, in which the Higgs basis coincides with the mass basis for the 125~GeV state, i.e.\ the field
\(H_0=\sum_a (v_a/v)\,H_a\) is itself a mass eigenstate (see e.g.\ \cite{Bento:2018fmy}). In the Higgs basis, exact alignment corresponds to the absence of mixing between the CP-even mode along the electroweak vev direction and the three orthogonal CP-even modes, i.e.\ \((M^2_{\rm even})_{0i}=0\) for \(i=1,2,3\): three algebraic relations among the scalar couplings. Since the extension by \(H_q\) introduces $5$ independent \(H_q\)--\(H_i\) quartic couplings (Eq.~\eqref{eq:potadd}), there is enough parameter freedom to satisfy the alignment conditions, and we have verified explicitly that solutions exist for general vevs \(v_q\) and \(v_i\). While the phenomenology of a general 4HDM can be considerably richer, it has been studied elsewhere and is not the focus of our work. Here we simply point out that the exact SM limit can be realised, in which the Higgs sector constraints are automatically satisfied; departures from this limit would translate into corresponding deviations from SM expectations in the scalar sector.

\bibliographystyle{utphys}
\bibliography{references.bib, biby.bib} 

@article{Bento:2018fmy,
    author = "Bento, Miguel P. and Haber, Howard E. and Rom{\~a}o, J. C. and Silva, Jo{\~a}o P.",
    title = "{Multi-Higgs doublet models: the Higgs-fermion couplings and their sum rules}",
    eprint = "1808.07123",
    archivePrefix = "arXiv",
    primaryClass = "hep-ph",
    reportNumber = "SCIPP-18/06",
    doi = "10.1007/JHEP10(2018)143",
    journal = "JHEP",
    volume = "10",
    pages = "143",
    year = "2018"
}

@article{Khan:2019doq,
    author = "Khan, Amir N. and Nunokawa, Hiroshi and Parke, Stephen J",
    title = "{Why matter effects matter for JUNO}",
    eprint = "1910.12900",
    archivePrefix = "arXiv",
    primaryClass = "hep-ph",
    reportNumber = "FERMILAB-PUB-19-490-T",
    doi = "10.1016/j.physletb.2020.135354",
    journal = "Phys. Lett. B",
    volume = "803",
    pages = "135354",
    year = "2020"
}

@article{JUNO:2025gmd,
    author = "Abusleme, Angel and others",
    collaboration = "JUNO",
    title = "{First measurement of reactor neutrino oscillations at JUNO}",
    eprint = "2511.14593",
    archivePrefix = "arXiv",
    primaryClass = "hep-ex",
    month = "11",
    year = "2025"
}

@inproceedings{Project8:2022wqh,
    author = "Esfahani, A. Ashtari and others",
    collaboration = "Project 8",
    title = "{The Project 8 Neutrino Mass Experiment}",
    booktitle = "{Snowmass 2021}",
    eprint = "2203.07349",
    archivePrefix = "arXiv",
    primaryClass = "nucl-ex",
    month = "3",
    year = "2022"
}

@article{Naredo-Tuero:2024sgf,
    author = "Naredo-Tuero, Daniel and Escudero, Miguel and Fern{\'a}ndez-Mart{\'\i}nez, Enrique and Marcano, Xabier and Poulin, Vivian",
    title = "{Critical look at the cosmological neutrino mass bound}",
    eprint = "2407.13831",
    archivePrefix = "arXiv",
    primaryClass = "astro-ph.CO",
    reportNumber = "CERN-TH-2024-115, IFT-UAM/CSIC-24-106",
    doi = "10.1103/PhysRevD.110.123537",
    journal = "Phys. Rev. D",
    volume = "110",
    number = "12",
    pages = "123537",
    year = "2024"
}

@article{Bertolez-Martinez:2024wez,
    author = "Bert{\'o}lez-Mart{\'\i}nez, Toni and Esteban, Ivan and Hajjar, Rasmi and Mena, Olga and Salvado, Jordi",
    title = "{Origin of cosmological neutrino mass bounds: background versus perturbations}",
    eprint = "2411.14524",
    archivePrefix = "arXiv",
    primaryClass = "astro-ph.CO",
    doi = "10.1088/1475-7516/2025/06/058",
    journal = "JCAP",
    volume = "06",
    pages = "058",
    year = "2025"
}

@article{Antusch:2007km,
    author = "Antusch, Stefan",
    title = "{Flavour-dependent type II leptogenesis}",
    eprint = "0704.1591",
    archivePrefix = "arXiv",
    primaryClass = "hep-ph",
    reportNumber = "FTUAM-07-07, IFT-UAM-CSIC-07-18",
    doi = "10.1103/PhysRevD.76.023512",
    journal = "Phys. Rev. D",
    volume = "76",
    pages = "023512",
    year = "2007"
}

@article{Antusch:2004xy,
    author = "Antusch, Stefan and King, Steve F.",
    title = "{Type II Leptogenesis and the neutrino mass scale}",
    eprint = "hep-ph/0405093",
    archivePrefix = "arXiv",
    reportNumber = "SHEP-0415",
    doi = "10.1016/j.physletb.2004.07.009",
    journal = "Phys. Lett. B",
    volume = "597",
    pages = "199--207",
    year = "2004"
}

@article{Borah:2018nvu, author = "Borah, Debasish and Karmakar, Biswajit", title = "{Linear seesaw for Dirac neutrinos with $A_4$ flavour symmetry}", eprint = "1806.10685", archivePrefix = "arXiv", primaryClass = "hep-ph", doi = "10.1016/j.physletb.2018.12.006", journal = "Phys. Lett. B", volume = "789", pages = "59--70", year = "2019" }

@article{Borah:2017dmk, author = "Borah, Debasish and Karmakar, Biswajit", title = "{$A_4$ flavour model for Dirac neutrinos: Type I and inverse seesaw}", eprint = "1712.06407", archivePrefix = "arXiv", primaryClass = "hep-ph", doi = "10.1016/j.physletb.2018.03.047", journal = "Phys. Lett. B", volume = "780", pages = "461--470", year = "2018" }

@article{Kumar:2023moh, author = "Kumar, Ranjeet and Mishra, Priya and Behera, Mitesh Kumar and Mohanta, Rukmani and Srivastava, Rahul", title = "{Predictions from scoto-seesaw with A4 modular symmetry}", eprint = "2310.02363", archivePrefix = "arXiv", primaryClass = "hep-ph", doi = "10.1016/j.physletb.2024.138635", journal = "Phys. Lett. B", volume = "853", pages = "138635", year = "2024" }

@article{CarcamoHernandez:2017kra, author = "C{\'a}rcamo Hern{\'a}ndez, A. E. and Long, H. N.", title = "{A highly predictive $A_{4}$ flavour 3-3-1 model with radiative inverse seesaw mechanism}", eprint = "1705.05246", archivePrefix = "arXiv", primaryClass = "hep-ph", doi = "10.1088/1361-6471/aaace7", journal = "J. Phys. G", volume = "45", number = "4", pages = "045001", year = "2018" }

@article{CarcamoHernandez:2015rmj, author = "C{\'a}rcamo Hern{\'a}ndez, A. E. and Martinez, Roberto", title = "{A predictive 3-3-1 model with $A_4$ flavor symmetry}", eprint = "1501.05937", archivePrefix = "arXiv", primaryClass = "hep-ph", doi = "10.1016/j.nuclphysb.2016.02.025", journal = "Nucl. Phys. B", volume = "905", pages = "337--358", year = "2016" }

@article{CarcamoHernandez:2013yiy, author = {Carcamo Hernandez, Antonio Enrique and de Medeiros Varzielas, Ivo and Kovalenko, S. G. and P{\"a}s, H. and Schmidt, Ivan}, title = "{Lepton masses and mixings in an $A_4$ multi-Higgs model with a radiative seesaw mechanism}", eprint = "1307.6499", archivePrefix = "arXiv", primaryClass = "hep-ph", doi = "10.1103/PhysRevD.88.076014", journal = "Phys. Rev. D", volume = "88", number = "7", pages = "076014", year = "2013" }

@article{Barrie:2021mwi,
    author = "Barrie, Neil D. and Han, Chengcheng and Murayama, Hitoshi",
    title = "{Affleck-Dine Leptogenesis from Higgs Inflation}",
    eprint = "2106.03381",
    archivePrefix = "arXiv",
    primaryClass = "hep-ph",
    doi = "10.1103/PhysRevLett.128.141801",
    journal = "Phys. Rev. Lett.",
    volume = "128",
    number = "14",
    pages = "141801",
    year = "2022"
}

@article{Barrie:2022cub,
    author = "Barrie, Neil D. and Han, Chengcheng and Murayama, Hitoshi",
    title = "{Type II Seesaw leptogenesis}",
    eprint = "2204.08202",
    archivePrefix = "arXiv",
    primaryClass = "hep-ph",
    reportNumber = "CTPU-PTC-22-04",
    doi = "10.1007/JHEP05(2022)160",
    journal = "JHEP",
    volume = "05",
    pages = "160",
    year = "2022"
}

@article{Ma:1998dx,
    author = "Ma, Ernest and Sarkar, Utpal",
    title = "{Neutrino masses and leptogenesis with heavy Higgs triplets}",
    eprint = "hep-ph/9802445",
    archivePrefix = "arXiv",
    reportNumber = "UCRHEP-T-209",
    doi = "10.1103/PhysRevLett.80.5716",
    journal = "Phys. Rev. Lett.",
    volume = "80",
    pages = "5716--5719",
    year = "1998"
}

@article{Moreno-Sanchez:2025bzz,
    author = "Moreno-S{\'a}nchez, Adri{\'a}n and Palavri{\'c}, Ajdin",
    title = "{Leptonic flavor from a modular A4 symmetry: UV mediators and SMEFT realizations}",
    eprint = "2505.01535",
    archivePrefix = "arXiv",
    primaryClass = "hep-ph",
    doi = "10.1103/hdgd-cq45",
    journal = "Phys. Rev. D",
    volume = "112",
    number = "7",
    pages = "075002",
    year = "2025"
}

@article{Pathak:2024sei,
    author = "Pathak, Gourab and Das, Pritam and Das, Mrinal Kumar",
    title = "{Neutrino mass genesis in scoto-inverse seesaw with modular $A_4$}",
    eprint = "2411.13895",
    archivePrefix = "arXiv",
    primaryClass = "hep-ph",
    doi = "10.1140/epjc/s10052-025-14263-1",
    journal = "Eur. Phys. J. C",
    volume = "85",
    number = "5",
    pages = "569",
    year = "2025"
}

@article{Palavric:2024gvu,
    author = "Palavri{\'c}, Ajdin",
    title = "{Discrete leptonic flavor symmetries: UV mediators and phenomenology}",
    eprint = "2408.16044",
    archivePrefix = "arXiv",
    primaryClass = "hep-ph",
    doi = "10.1103/PhysRevD.110.115025",
    journal = "Phys. Rev. D",
    volume = "110",
    number = "11",
    pages = "115025",
    year = "2024"
}

@article{Denton:2023hkx,
    author = "Denton, Peter B. and Gehrlein, Julia",
    title = "{Survey of neutrino flavor predictions and the neutrinoless double beta decay funnel}",
    eprint = "2308.09737",
    archivePrefix = "arXiv",
    primaryClass = "hep-ph",
    reportNumber = "CERN-TH-2023-160, CETUP-2023-006",
    doi = "10.1103/PhysRevD.109.055028",
    journal = "Phys. Rev. D",
    volume = "109",
    number = "5",
    pages = "055028",
    year = "2024"
}

@article{Berbig:2025hlc,
    author = "Berbig, Maximilian",
    title = "{Type II Seesaw Leptogenesis in a Majoron background}",
    eprint = "2506.23290",
    archivePrefix = "arXiv",
    primaryClass = "hep-ph",
    month = "6",
    year = "2025"
}

@article{Nomura:2024atp,
    author = "Nomura, Takaaki and Okada, Hiroshi",
    title = "{Type-II seesaw of a non-holomorphic modular A4 symmetry}",
    eprint = "2408.01143",
    archivePrefix = "arXiv",
    primaryClass = "hep-ph",
    doi = "10.1016/j.physletb.2025.139763",
    journal = "Phys. Lett. B",
    volume = "868",
    pages = "139763",
    year = "2025"
}

@article{Planck:2018vyg,
    author = "Aghanim, N. and others",
    collaboration = "Planck",
    title = "{Planck 2018 results. VI. Cosmological parameters}",
    eprint = "1807.06209",
    archivePrefix = "arXiv",
    primaryClass = "astro-ph.CO",
    doi = "10.1051/0004-6361/201833910",
    journal = "Astron. Astrophys.",
    volume = "641",
    pages = "A6",
    year = "2020",
    note = "[Erratum: Astron.Astrophys. 652, C4 (2021)]"
}

@article{ParticleDataGroup:2024cfk,
    author = "Navas, S. and others",
    collaboration = "Particle Data Group",
    title = "{Review of particle physics}",
    doi = "10.1103/PhysRevD.110.030001",
    journal = "Phys. Rev. D",
    volume = "110",
    number = "3",
    pages = "030001",
    year = "2024"
}

@article{Kajita:2016cak,
    author = "Kajita, Takaaki",
    title = "{Nobel Lecture: Discovery of atmospheric neutrino oscillations}",
    doi = "10.1103/RevModPhys.88.030501",
    journal = "Rev. Mod. Phys.",
    volume = "88",
    number = "3",
    pages = "030501",
    year = "2016"
}

@article{Schechter:1980gr,
    author = "Schechter, J. and Valle, J. W. F.",
    title = "{Neutrino Masses in SU(2) x U(1) Theories}",
    reportNumber = "SU-4217-167, COO-3533-167",
    doi = "10.1103/PhysRevD.22.2227",
    journal = "Phys. Rev. D",
    volume = "22",
    pages = "2227",
    year = "1980"
}

@article{Mahapatra:2023oyh,
    author = "Mahapatra, Satyabrata and Sahoo, Sujit Kumar and Sahu, Narendra and Thounaojam, Vicky Singh",
    title = "{Self-interacting dark matter and Dirac neutrinos via lepton quarticity}",
    eprint = "2312.12322",
    archivePrefix = "arXiv",
    primaryClass = "hep-ph",
    doi = "10.1103/PhysRevD.109.055036",
    journal = "Phys. Rev. D",
    volume = "109",
    number = "5",
    pages = "055036",
    year = "2024"
}

@article{CentellesChulia:2017koy,
        author = "Centelles Chuli{\'a}, Salvador and Srivastava, Rahul and Valle, Jos{\'e} W. F.",
         title = "{Generalized Bottom-Tau unification, neutrino oscillations and dark matter: predictions from a lepton quarticity flavor approach}",
       journal = "Phys.Lett.",
        volume = "B773",
          year = "2017",
         pages = "26-33",
 archiveprefix = "arXiv",
  primaryclass = "hep-ph",
        eprint = "1706.00210",
           doi = "10.1016/j.physletb.2017.07.065",
}

@article{Bento:2023owf,
    author = "Bento, Miguel P. and Silva, Joao P. and Trautner, Andreas",
    title = "{The basis invariant flavor puzzle}",
    eprint = "2308.00019",
    archivePrefix = "arXiv",
    primaryClass = "hep-ph",
    doi = "10.1007/JHEP01(2024)024",
    journal = "JHEP",
    volume = "01",
    pages = "024",
    year = "2024"
}

@article{Nakane:2025yxq,
    author = "Nakane, Jun",
    title = "{New detector development and performance evaluation for KamLAND2-Zen experiment}",
    doi = "10.1016/j.nima.2025.170850",
    journal = "Nucl. Instrum. Meth. A",
    volume = "1081",
    pages = "170850",
    year = "2026"
}

@article{Ding:2024ozt,
    author = "Ding, Gui-Jun and Valle, Jose W. F.",
    title = "{The symmetry approach to quark and lepton masses and mixing}",
    eprint = "2402.16963",
    archivePrefix = "arXiv",
    primaryClass = "hep-ph",
    doi = "10.1016/j.physrep.2024.12.005",
    journal = "Phys. Rept.",
    volume = "1109",
    pages = "1--105",
    year = "2025"
}

@article{Morisi:2012fg,
        author = "Morisi, S. and Valle, J.~W.~F.",
         title = "{Neutrino masses and mixing: a flavour symmetry roadmap}",
       journal = "Fortsch.Phys.",
        volume = "61",
          year = "2013",
         pages = "466-492",
 archiveprefix = "arXiv",
  primaryclass = "hep-ph",
        eprint = "1206.6678",
           doi = "10.1002/prop.201200125",
  reportnumber = "IFIC-12-46",
}

@article{Rodejohann:2011vc,
        author = "Rodejohann, W. and Valle, J.~W.~F.",
         title = "{Symmetrical Parametrizations of the Lepton Mixing Matrix}",
       journal = "Phys.Rev.",
        volume = "D84",
          year = "2011",
         pages = "073011",
 archiveprefix = "arXiv",
  primaryclass = "hep-ph",
        eprint = "1108.3484",
           doi = "10.1103/PhysRevD.84.073011",
  reportnumber = "IFIC-11-39",
}

@article{Babu:2002dz,
        author = "Babu, K.S. and Ma, Ernest and Valle, J.~W.~F.",
         title = "{Underlying A(4) symmetry for the neutrino mass matrix and the quark mixing matrix}",
       journal = "Phys.Lett.",
        volume = "B552",
          year = "2003",
         pages = "207-213",
           doi = "10.1016/S0370-2693(02)03153-2",
         arxiv = "hep-ph/0206292",
  reportnumber = "IFIC-02-26",
}

@article{McDonald:2016ixn,
        author = "McDonald, Arthur B.",
         title = "{Nobel Lecture: The Sudbury Neutrino Observatory: Observation of flavor change for solar neutrinos}",
       journal = "Rev.Mod.Phys.",
        volume = "88",
          year = "2016",
         pages = "030502",
           doi = "10.1103/RevModPhys.88.030502",
}

@article{Ding:2020vud,
    author = "Ding, Gui-Jun and Lu, Jun-Nan and Valle, Jos\'e W. F.",
    title = "{Trimaximal neutrino mixing from scotogenic $A_4$ family symmetry}",
    eprint = "2009.04750",
    archivePrefix = "arXiv",
    primaryClass = "hep-ph",
    reportNumber = "USTC-ICTS/PCFT-20-27, IFIC/20-XXX",
    doi = "10.1016/j.physletb.2021.136122",
    journal = "Phys. Lett. B",
    volume = "815",
    pages = "136122",
    year = "2021"
}

@article{Hyper-Kamiokande:2018ofw,
        author = "Abe, K. and others",
 collaboration = "Hyper-Kamiokande",
         title = "{Hyper-Kamiokande Design Report}",
          year = "2018",
 archiveprefix = "arXiv",
  primaryclass = "physics.ins-det",
        eprint = "1805.04163",
         month = "5",
}

@article{Ishimori:2010au,
        author = "Ishimori, Hajime and others",
         title = "{Non-Abelian Discrete Symmetries in Particle Physics}",
       journal = "Prog.Theor.Phys.Suppl.",
        volume = "183",
          year = "2010",
         pages = "1-163",
 archiveprefix = "arXiv",
  primaryclass = "hep-th",
        eprint = "1003.3552",
           doi = "10.1143/PTPS.183.1",
  reportnumber = "KUNS-2260",
}

@article{Feruglio:2019ybq,
        author = "Feruglio, Ferruccio and Romanino, Andrea",
         title = "{Lepton flavor symmetries}",
       journal = "Rev.Mod.Phys.",
        volume = "93",
          year = "2021",
         pages = "015007",
 archiveprefix = "arXiv",
  primaryclass = "hep-ph",
        eprint = "1912.06028",
           doi = "10.1103/RevModPhys.93.015007",
}

@article{Magg:1980ut,
    author = "Magg, M. and Wetterich, C.",
    title = "{Neutrino Mass Problem and Gauge Hierarchy}",
    reportNumber = "CERN-TH-2829",
    doi = "10.1016/0370-2693(80)90825-4",
    journal = "Phys. Lett. B",
    volume = "94",
    pages = "61--64",
    year = "1980"
}

@article{Mohapatra:1980yp,
    author = "Mohapatra, Rabindra N. and Senjanovic, Goran",
    title = "{Neutrino Masses and Mixings in Gauge Models with Spontaneous Parity Violation}",
    reportNumber = "FERMILAB-PUB-80-061-THY, FERMILAB-PUB-80-061-T",
    doi = "10.1103/PhysRevD.23.165",
    journal = "Phys. Rev. D",
    volume = "23",
    pages = "165",
    year = "1981"
}

@article{Cheng:1980qt,
    author = "Cheng, T. P. and Li, Ling-Fong",
    title = "{Neutrino Masses, Mixings and Oscillations in SU(2) x U(1) Models of Electroweak Interactions}",
    reportNumber = "PRINT-80-0511 (CARNEGIE-MELLON), COO-3066-152",
    doi = "10.1103/PhysRevD.22.2860",
    journal = "Phys. Rev. D",
    volume = "22",
    pages = "2860",
    year = "1980"
}

@article{deMedeirosVarzielas:2025byb,
    author = "de Medeiros Varzielas, Ivo and Liu, Ming-Shau and Sengupta, Amartya and Talbert, Jim",
    title = "{Residual Symmetries and Scalar Multiplet Vacuum Alignment in Non-Abelian Flavour Models}",
    eprint = "2512.19789",
    archivePrefix = "arXiv",
    primaryClass = "hep-ph",
    reportNumber = "LA-UR-25-31854",
    month = "12",
    year = "2025"
}

@article{Singh:2024imk,
    author = "Singh, Labh and Kashav, Monal and Verma, Surender",
    title = "{Minimal type-I Dirac seesaw and leptogenesis under A4 modular invariance}",
    eprint = "2405.07165",
    archivePrefix = "arXiv",
    primaryClass = "hep-ph",
    doi = "10.1016/j.nuclphysb.2024.116666",
    journal = "Nucl. Phys. B",
    volume = "1007",
    pages = "116666",
    year = "2024"
}

@article{Altarelli:2005yx,
    author = "Altarelli, Guido and Feruglio, Ferruccio",
    title = "{Tri-bimaximal neutrino mixing, A(4) and the modular symmetry}",
    eprint = "hep-ph/0512103",
    archivePrefix = "arXiv",
    reportNumber = "CERN-PH-TH-2005-226",
    doi = "10.1016/j.nuclphysb.2006.02.015",
    journal = "Nucl. Phys. B",
    volume = "741",
    pages = "215--235",
    year = "2006"
}

@article{T2K:2023smv,
    author = "Abe, K. and others",
    collaboration = "T2K",
    title = "{Measurements of neutrino oscillation parameters from the T2K experiment using $3.6\times 10^{21}$ protons on target}",
    eprint = "2303.03222",
    archivePrefix = "arXiv",
    primaryClass = "hep-ex",
    doi = "10.1140/epjc/s10052-023-11819-x",
    journal = "Eur. Phys. J. C",
    volume = "83",
    number = "9",
    pages = "782",
    year = "2023"
}

@article{Gehrlein:2020jnr,
    author = "Gehrlein, J. and Spinrath, M.",
    title = "{Leptonic Sum Rules from Flavour Models with Modular Symmetries}",
    eprint = "2012.04131",
    archivePrefix = "arXiv",
    primaryClass = "hep-ph",
    doi = "10.1007/JHEP03(2021)177",
    journal = "JHEP",
    volume = "03",
    pages = "177",
    year = "2021"
}

@article{King:2013psa,
    author = "King, Stephen F. and Merle, Alexander and Stuart, Alexander J.",
    title = "{The Power of Neutrino Mass Sum Rules for Neutrinoless Double Beta Decay Experiments}",
    eprint = "1307.2901",
    archivePrefix = "arXiv",
    primaryClass = "hep-ph",
    doi = "10.1007/JHEP12(2013)005",
    journal = "JHEP",
    volume = "12",
    pages = "005",
    year = "2013"
}

@article{Gehrlein:2016wlc,
    author = "Gehrlein, Julia and Merle, Alexander and Spinrath, Martin",
    title = "{Predictivity of Neutrino Mass Sum Rules}",
    eprint = "1606.04965",
    archivePrefix = "arXiv",
    primaryClass = "hep-ph",
    reportNumber = "MPP-2016-127, TTP16-022",
    doi = "10.1103/PhysRevD.94.093003",
    journal = "Phys. Rev. D",
    volume = "94",
    number = "9",
    pages = "093003",
    year = "2016"
}

@article{NOvA:2021nfi,
    author = "Acero, M. A. and others",
    collaboration = "NOvA",
    title = "{Improved measurement of neutrino oscillation parameters by the NOvA experiment}",
    eprint = "2108.08219",
    archivePrefix = "arXiv",
    primaryClass = "hep-ex",
    reportNumber = "FERMILAB-PUB-21-373-ND",
    doi = "10.1103/PhysRevD.106.032004",
    journal = "Phys. Rev. D",
    volume = "106",
    number = "3",
    pages = "032004",
    year = "2022"
}

@article{DUNE:2020ypp,
    author = "Abi, Babak and others",
    collaboration = "DUNE",
    title = "{Deep Underground Neutrino Experiment (DUNE), Far Detector Technical Design Report, Volume II: DUNE Physics}",
    eprint = "2002.03005",
    archivePrefix = "arXiv",
    primaryClass = "hep-ex",
    reportNumber = "FERMILAB-PUB-20-025-ND, FERMILAB-DESIGN-2020-02",
    month = "2",
    year = "2020"
}

@article{Barry:2010yk,
    author = "Barry, James and Rodejohann, Werner",
    title = "{Neutrino Mass Sum-rules in Flavor Symmetry Models}",
    eprint = "1007.5217",
    archivePrefix = "arXiv",
    primaryClass = "hep-ph",
    doi = "10.1016/j.nuclphysb.2010.08.015",
    journal = "Nucl. Phys. B",
    volume = "842",
    pages = "33--50",
    year = "2011"
}

@article{deSalas:2020pgw,
    author = "de Salas, P. F. and Forero, D. V. and Gariazzo, S. and Mart\'\i{}nez-Mirav\'e, P. and Mena, O. and Ternes, C. A. and T\'ortola, M. and Valle, J. W. F.",
    title = "{2020 global reassessment of the neutrino oscillation picture}",
    eprint = "2006.11237",
    archivePrefix = "arXiv",
    primaryClass = "hep-ph",
    doi = "10.1007/JHEP02(2021)071",
    journal = "JHEP",
    volume = "02",
    pages = "071",
    year = "2021"
}

@article{KamLAND-Zen:2024eml,
    author = "Abe, S. and others",
    collaboration = "KamLAND-Zen",
    title = "{Search for Majorana Neutrinos with the Complete KamLAND-Zen Dataset}",
    eprint = "2406.11438",
    archivePrefix = "arXiv",
    primaryClass = "hep-ex",
    doi = "10.1103/jkf6-48j8",
    journal = "Phys. Rev. Lett.",
    volume = "135",
    number = "26",
    pages = "262501",
    year = "2025"
}

@article{Calibbi:2025fzi,
    author = "Calibbi, Lorenzo and Hagedorn, Claudia and Schmidt, Michael A. and Vandeleur, James",
    title = "{Selection rules for charged lepton flavor violating processes from residual flavor groups}",
    eprint = "2505.24350",
    archivePrefix = "arXiv",
    primaryClass = "hep-ph",
    doi = "10.1103/b6gz-c1c9",
    journal = "Phys. Rev. D",
    volume = "112",
    number = "7",
    pages = "075031",
    year = "2025"
}

@article{Goswami:2025jde,
    author = "Goswami, Sagar Tirtha and Roy, Subhankar",
    title = "{Permuted charged lepton correction in the framework of dirac seesaw}",
    eprint = "2501.18181",
    archivePrefix = "arXiv",
    primaryClass = "hep-ph",
    doi = "10.1016/j.nuclphysb.2025.117275",
    journal = "Nucl. Phys. B",
    volume = "1022",
    pages = "117275",
    year = "2026"
}

@article{Kumar:2025zvv,
    author = "Kumar, Ranjeet and Prajapati, Hemant Kumar and Srivastava, Rahul and Yadav, Sushant",
    title = "{Flavor imprints on novel low mass dark matter}",
    eprint = "2510.02972",
    archivePrefix = "arXiv",
    primaryClass = "hep-ph",
    doi = "10.1007/JHEP11(2025)094",
    journal = "JHEP",
    volume = "11",
    pages = "094",
    year = "2025"
}

@article{Kumar:2025cte,
    author = "Kumar, Ranjeet and Nath, Newton and Srivastava, Rahul and Yadav, Sushant",
    title = "{Dirac Scoto inverse-seesaw from A$_{4}$ flavor symmetry}",
    eprint = "2505.01407",
    archivePrefix = "arXiv",
    primaryClass = "hep-ph",
    doi = "10.1007/JHEP10(2025)088",
    journal = "JHEP",
    volume = "10",
    pages = "088",
    year = "2025"
}

@article{Kumar:2024zfb,
    author = "Kumar, Ranjeet and Nath, Newton and Srivastava, Rahul",
    title = "{Cutting the scotogenic loop: adding flavor to dark matter}",
    eprint = "2406.00188",
    archivePrefix = "arXiv",
    primaryClass = "hep-ph",
    doi = "10.1007/JHEP12(2024)036",
    journal = "JHEP",
    volume = "12",
    pages = "036",
    year = "2024"
}

@article{RickyDevi:2023fqd,
    author = "Ricky Devi, Maibam and Bora, Kalpana",
    title = "{Linking resonant leptogenesis with dynamics of the inverse seesaw theory with $ A_{4} $ flavor symmetry}",
    eprint = "2304.13546",
    archivePrefix = "arXiv",
    primaryClass = "hep-ph",
    month = "4",
    year = "2023"
}

@article{Bora:2022jdp,
    author = "Bora, Kalpana and Devi, Maibam Ricky",
    title = "{Exploring Dynamics of A$_{4}$ Flavour Symmetry Using Low Scale Seesaw Mechanisms}",
    eprint = "2212.11492",
    archivePrefix = "arXiv",
    primaryClass = "hep-ph",
    doi = "10.1134/S154747712206005X",
    journal = "Phys. Part. Nucl. Lett.",
    volume = "19",
    number = "6",
    pages = "642--645",
    year = "2022"
}

@article{Devi:2022scm,
    author = "Devi, Maibam Ricky and Bora, Kalpana",
    title = "{Exploring the feasibility of the charged lepton flavor violating decay {\ensuremath{\mu}} {\textrightarrow} e + {\ensuremath{\gamma}} in inverse and linear seesaw mechanisms with A4 flavor symmetry}",
    eprint = "2208.02214",
    archivePrefix = "arXiv",
    primaryClass = "hep-ph",
    doi = "10.1142/S0217732322502066",
    journal = "Mod. Phys. Lett. A",
    volume = "37",
    number = "31",
    pages = "2250206",
    year = "2022"
}

@article{Project8:2025aar,
    author = "Ashtari Esfahani, A. and others",
    collaboration = "Project 8",
    title = "{Antenna arrays for neutrino mass measurements with cyclotron radiation emission spectroscopy}",
    doi = "10.1103/dc29-s4y2",
    journal = "Phys. Rev. C",
    volume = "112",
    number = "4",
    pages = "045506",
    year = "2025"
}

@article{Project8:2022hun,
    author = "Ashtari Esfahani, A. and others",
    collaboration = "Project 8",
    title = "{Tritium Beta Spectrum Measurement and Neutrino Mass Limit from Cyclotron Radiation Emission Spectroscopy}",
    eprint = "2212.05048",
    archivePrefix = "arXiv",
    primaryClass = "nucl-ex",
    doi = "10.1103/PhysRevLett.131.102502",
    journal = "Phys. Rev. Lett.",
    volume = "131",
    number = "10",
    pages = "102502",
    year = "2023"
}

@article{KATRIN:2024cdt,
    author = "Aker, Max and others",
    collaboration = "KATRIN",
    title = "{Direct neutrino-mass measurement based on 259 days of KATRIN data}",
    eprint = "2406.13516",
    archivePrefix = "arXiv",
    primaryClass = "nucl-ex",
    doi = "10.1126/science.adq9592",
    journal = "Science",
    volume = "388",
    number = "6743",
    pages = "adq9592",
    year = "2025"
}

@article{Amendola:2016saw,
    author = "Amendola, Luca and others",
    title = "{Cosmology and fundamental physics with the Euclid satellite}",
    eprint = "1606.00180",
    archivePrefix = "arXiv",
    primaryClass = "astro-ph.CO",
    doi = "10.1007/s41114-017-0010-3",
    journal = "Living Rev. Rel.",
    volume = "21",
    number = "1",
    pages = "2",
    year = "2018"
}

@article{SPT-3G:2025bzu,
    author = "Camphuis, E. and others",
    collaboration = "SPT-3G",
    title = "{SPT-3G D1: CMB temperature and polarization power spectra and cosmology from 2019 and 2020 observations of the SPT-3G Main field}",
    eprint = "2506.20707",
    archivePrefix = "arXiv",
    primaryClass = "astro-ph.CO",
    reportNumber = "FERMILAB-PUB-25-0144-PPD",
    month = "6",
    year = "2025"
}

@article{Chen:2005jm,
    author = "Chen, Shao-Long and Frigerio, Michele and Ma, Ernest",
    title = "{Hybrid seesaw neutrino masses with A(4) family symmetry}",
    eprint = "hep-ph/0504181",
    archivePrefix = "arXiv",
    reportNumber = "UCRHEP-T387",
    doi = "10.1016/j.nuclphysb.2005.07.012",
    journal = "Nucl. Phys. B",
    volume = "724",
    pages = "423--431",
    year = "2005"
}

@article{Babu:1990fr,
    author = "Babu, K. S. and Mohapatra, R. N.",
    title = "{Permutation Symmetry and the Origin of Fermion Mass Hierarchy}",
    reportNumber = "MdDP-PP-90-158",
    doi = "10.1103/PhysRevLett.64.2747",
    journal = "Phys. Rev. Lett.",
    volume = "64",
    pages = "2747",
    year = "1990"
}

@article{Frampton:1994rk,
    author = "Frampton, P. H. and Kephart, T. W.",
    title = "{Simple nonAbelian finite flavor groups and fermion masses}",
    eprint = "hep-ph/9409330",
    archivePrefix = "arXiv",
    reportNumber = "IFP-702-UNC, VAND-TH-94-8",
    doi = "10.1142/S0217751X95002187",
    journal = "Int. J. Mod. Phys. A",
    volume = "10",
    pages = "4689--4704",
    year = "1995"
}

@article{Kubo:2003iw,
    author = "Kubo, J. and Mondragon, A. and Mondragon, M. and Rodriguez-Jauregui, E.",
    title = "{The Flavor symmetry}",
    eprint = "hep-ph/0302196",
    archivePrefix = "arXiv",
    reportNumber = "KANAZAWA-03-05",
    doi = "10.1143/PTP.109.795",
    journal = "Prog. Theor. Phys.",
    volume = "109",
    pages = "795--807",
    year = "2003",
    note = "[Erratum: Prog.Theor.Phys. 114, 287--287 (2005)]"
}

@article{Altarelli:2010gt,
    author = "Altarelli, Guido and Feruglio, Ferruccio",
    title = "{Discrete Flavor Symmetries and Models of Neutrino Mixing}",
    eprint = "1002.0211",
    archivePrefix = "arXiv",
    primaryClass = "hep-ph",
    reportNumber = "RM3-TH-10-01, CERN-PH-TH-2010-016, DFPD-10-TH-02",
    doi = "10.1103/RevModPhys.82.2701",
    journal = "Rev. Mod. Phys.",
    volume = "82",
    pages = "2701--2729",
    year = "2010"
}

@article{Ma:2010gs,
    author = "Ma, Ernest",
    title = "{Quark and Lepton Flavor Triality}",
    eprint = "1006.3524",
    archivePrefix = "arXiv",
    primaryClass = "hep-ph",
    reportNumber = "UCRHEP-T492",
    doi = "10.1103/PhysRevD.82.037301",
    journal = "Phys. Rev. D",
    volume = "82",
    pages = "037301",
    year = "2010"
}

@article{Ma:2001dn,
    author = "Ma, Ernest and Rajasekaran, G.",
    title = "{Softly broken A(4) symmetry for nearly degenerate neutrino masses}",
    eprint = "hep-ph/0106291",
    archivePrefix = "arXiv",
    reportNumber = "UCRHEP-T308",
    doi = "10.1103/PhysRevD.64.113012",
    journal = "Phys. Rev. D",
    volume = "64",
    pages = "113012",
    year = "2001"
}

@article{CentellesChulia:2023osj,
    author = "Centelles Chuli{\'a}, Salvador and Kumar, Ranjeet and Popov, Oleg and Srivastava, Rahul",
    title = "{Neutrino mass sum rules from modular A4 symmetry}",
    eprint = "2308.08981",
    archivePrefix = "arXiv",
    primaryClass = "hep-ph",
    doi = "10.1103/PhysRevD.109.035016",
    journal = "Phys. Rev. D",
    volume = "109",
    number = "3",
    pages = "035016",
    year = "2024"
}

@article{ChuliaCentelles:2022ogm,
    author = "Chuli{\'a} Centelles, Salvador and Cepedello, Ricardo and Medina, Omar",
    title = "{Absolute neutrino mass scale and dark matter stability from flavour symmetry}",
    eprint = "2204.12517",
    archivePrefix = "arXiv",
    primaryClass = "hep-ph",
    doi = "10.1007/JHEP10(2022)080",
    journal = "JHEP",
    volume = "10",
    pages = "080",
    year = "2022"
}

@article{Cao:2011df,
    author = "Cao, Qing-Hong and Damanik, Asan and Ma, Ernest and Wegman, Daniel",
    title = "{Probing Lepton Flavor Triality with Higgs Boson Decay}",
    eprint = "1103.0008",
    archivePrefix = "arXiv",
    primaryClass = "hep-ph",
    reportNumber = "ANL-HEP-PR-11-14, EFI-11-6, UCRHEP-T500",
    doi = "10.1103/PhysRevD.83.093012",
    journal = "Phys. Rev. D",
    volume = "83",
    pages = "093012",
    year = "2011"
}

@article{Mandal:2022zmy,
    author = "Mandal, Sanjoy and Miranda, O. G. and Sanchez Garcia, G. and Valle, J. W. F. and Xu, Xun-Jie",
    title = "{Toward deconstructing the simplest seesaw mechanism}",
    eprint = "2203.06362",
    archivePrefix = "arXiv",
    primaryClass = "hep-ph",
    doi = "10.1103/PhysRevD.105.095020",
    journal = "Phys. Rev. D",
    volume = "105",
    number = "9",
    pages = "095020",
    year = "2022"
}

@article{Buskin:2021eig,
    author = "Buskin, N. and Ivanov, Igor P.",
    title = "{Bounded-from-below conditions for $A_4$-symmetric 3HDM}",
    eprint = "2104.11428",
    archivePrefix = "arXiv",
    primaryClass = "hep-ph",
    doi = "10.1088/1751-8121/ac0e53",
    journal = "J. Phys. A",
    volume = "54",
    pages = "325401",
    year = "2021"
}

@article{Lavoura:2003xp,
    author = "Lavoura, L.",
    title = "{General formulae for f(1) ---{\ensuremath{>}} f(2) gamma}",
    eprint = "hep-ph/0302221",
    archivePrefix = "arXiv",
    doi = "10.1140/epjc/s2003-01212-7",
    journal = "Eur. Phys. J. C",
    volume = "29",
    pages = "191--195",
    year = "2003"
}

@article{Barrie:2022ake,
    author = "Barrie, N. D. and Petcov, S. T.",
    title = "{Lepton Flavour Violation tests of Type II Seesaw Leptogenesis}",
    eprint = "2210.02110",
    archivePrefix = "arXiv",
    primaryClass = "hep-ph",
    reportNumber = "CTPU-PTC-22-18, SISSA 17/2022/FSI",
    doi = "10.1007/JHEP01(2023)001",
    journal = "JHEP",
    volume = "01",
    pages = "001",
    year = "2023"
}

@article{Dinh:2012bp,
    author = "Dinh, D. N. and Ibarra, A. and Molinaro, E. and Petcov, S. T.",
    title = "{The $\mu - e$ Conversion in Nuclei, $\mu \to e \gamma, \mu \to 3e$ Decays and TeV Scale See-Saw Scenarios of Neutrino Mass Generation}",
    eprint = "1205.4671",
    archivePrefix = "arXiv",
    primaryClass = "hep-ph",
    reportNumber = "FLAVOUR(267104)-ERC-15, SISSA-10-2012-EP, TUM-HEP-837-12, CFTP-12-007",
    doi = "10.1007/JHEP08(2012)125",
    journal = "JHEP",
    volume = "08",
    pages = "125",
    year = "2012",
    note = "[Erratum: JHEP 09, 023 (2013)]"
}
\end{document}